\documentclass[12pt]{article}

\def\be{\begin{equation}}
\def\ee{\end{equation}}
\def\bea{\begin{eqnarray}}
\def\eea{\end{eqnarray}}
\def\beq{\begin{eqnarray}}
\def\eeq{\end{eqnarray}}
\def\eval#1{\left\langle#1\right\rangle}
\def\bA{{\bf A}}
\def\bm{{\bf m}}
\def\bn{{\bf n}}

\def\bk{{\bf k}}
\def\bx{{\bf x}}
\def\by{{\bf y}}

\usepackage{amssymb,amsmath}
\usepackage{amsfonts}
\usepackage{epsf}

\newcommand{\tr}{{\mathrm{tr}}}
\newcommand{\TR}{{\mathrm{Tr}}}

\begin{document}
\begin{titlepage}

\centerline{{\Large \bf Variational techniques in non-perturbative QCD }}
\vskip 0.7cm
\centerline{\large Alex Kovner ${}^{1,2}$ and J. Guilherme Milhano ${}^{3,4}$}

\vskip 0.3cm

\centerline{${}^1$School of Mathematics and Statistics, University of Plymouth,}
\centerline{Drake Circus, PL4 8AA, United Kingdom}
\vskip 0.2cm
\centerline{${}^2$Physics Department, University of Connecticut, 2152 Hillside Road,}
\centerline{Storrs, CT 06269-3046, USA}
\vskip 0.2cm
\centerline{${}^3$ CENTRA, Instituto Superior T\'ecnico (IST),}
\centerline{Av. Rovisco Pais, P-1049-001 Lisboa, Portugal}
\vskip 0.2cm
\centerline{${}^4$ Department of Physics}
\centerline{CERN, CH-1211 Geneva 23, Switzerland}
\vskip 0.2cm

\vskip 0.2cm
\begin{center}
\noindent
{\small\em To be published in the Ian Kogan Memorial Volume ``From Fields to Strings: Circumnavigating Theoretical Physics",  World Scientific, 2004}
\end{center}

\vskip 0.7cm

%

\abstract{We review attempts to apply the variational
principle to understand the vacuum of non-abelian gauge
theories. In particular, we focus on the method explored by Ian
Kogan and collaborators, which imposes exact gauge invariance on
the trial Gaussian wave functional prior to the minimization of
energy. We describe the application of the method to a toy model ---
confining compact QED in 2+1 dimensions --- where it works
wonderfully and reproduces all known non-trivial results. We then
follow its applications to pure Yang-Mills theory in 3+1
dimensions at zero and finite temperature. Among the results of
the variational calculation are dynamical mass generation and the
analytic description of the deconfinement phase transition.}

\end{titlepage}


\newpage

\tableofcontents
\newpage

\section{Introduction}

Quantum Chromodynamics (QCD) was formulated more than 30 years
ago. There is very little doubt that it is, indeed, the correct
theory of strong interactions. An impressive amount of
experimental data is successfully described by QCD calculations in
the high momentum transfer domain. For these processes it is the
short time - short distance dynamics that is relevant.
Fortunately, in this domain QCD is weakly coupled and thus precise
quantitative information can be obtained by analytic calculations.

Of course, high momentum transfer processes are not the only, and perhaps not the most interesting,
domain governed by QCD dynamics.
The vacuum --- low energy, large distance --- sector of the strong interactions exhibits a wealth of phenomena which are both qualitatively  striking at the experimental level
 and notoriously difficult to establish within the fundamental framework of QCD.
Understanding low energy phenomena in QCD, such as confinement and
chiral symmetry breaking or, in more general terms, the strong
coupling problem  and  the ground state structure of an
asymptotically free non-abelian gauge theory is, without doubt,
one of  the main problems in the modern quantum field theory. In
spite of years of attempts to answer these questions we are still
far from achieving this goal.

Many routes have been tried in approaching this problem. They
range from the highly computer intensive numerical programme of
the lattice gauge theory \cite{lattice}, through the universality
based concepts of  effective field theories \cite{effective} to
the monopole and vortex inspired searches for the confinement
mechanism \cite{monopolevortex,mongreen}. In recent years, there have also
been attempts to approach the non-perturbative physics of QCD
utilizing the information about supersymmetric gauge theories
\cite{susy}.

While all these ideas are interesting and productive, each has its
own drawbacks. The lattice gauge theory aims, in principle, at
producing the complete set of numbers that characterize the  QCD
spectrum, condensates, various matrix elements, etc. This goal is,
however, still not within reach. Also often one would like to
understand the underlying physics rather than just calculate a
given number albeit with an accuracy of a fraction of a percent,
and this is difficult within a numerical approach. The effective
field theory and dual superconductor approaches are based on
simple physical pictures, but have their starting points rather
far from actual QCD, so that making quantitative contact with QCD
dynamics is very difficult. The SUSY motivated route suffers from
the same basic problem, as it is not in principle clear how much
the QCD dynamics is distorted by supersymmetry.

One should therefore welcome any method which attempts to analytically obtain dynamical information
directly from QCD, even if this information may be partial and incomplete.
Such a method should be, of course, intrinsically non-perturbative.
Unfortunately, the arsenal of non-perturbative methods to
 tackle strongly interacting quantum field theories (QFT) is, to say the least, very limited.
Methods that perform very well in simple quantum mechanical problems
are much more difficult to use in QFT.
This is true,
for example, for
a variational approach. In quantum mechanics it is usually enough
to know a few simple qualitative features in order to set up a
variational Ansatz which gives
pretty accurate results, not only for the energy of a ground
state, but also
for various other vacuum expectation values (VEV).
In QFT one is immediately faced with several difficult
problems when trying to apply this method, as discussed
insightfully by  Feynman \cite{vangerooge}.

Nevertheless over the years there have been several attempts to
apply different versions of the variational Rayleigh-Ritz method
to QCD. The purpose of this article is to review the results of
this approach. We will concentrate most of our attention on the
recent incarnation of the variational method pioneered by Ian
Kogan with collaborators \cite{Kogan:1995wf}, which in terms of
applications and results made much more headway than any of the
previous attempts. This is not to say that it is free from
problems and immune to criticism. Still, we believe that this is a
good time to summarize its status for two reasons. First because
some interesting quantitative results have been already achieved;
and, second, because there is plenty of room for improving the
method, and some aspects can be improved with relative ease, so
that clearly the method has not yet outlived its usefulness.

As mentioned earlier, when applying a variational method to QFT one is faced with many difficulties.
First of all there is a problem of generality of a trial state. The trial state ought to
be general enough to allow for, through variation of its parameters, the relevant physics to be spanned.
In quantum mechanics the task can be put down simply to identifying a few 
critical physical properties and consequently writing a
compliant trial state.
On the other hand, the Hilbert space of QFT is enormous, and it is much more difficult to identify ``by pure thought''
the relevant characteristics that have to be probed.

Then there is the problem of calculability.
That is, even if one
had a very good guess at the form of the vacuum wave functional (or,
for that matter,  even knew its exact form) one would still have to
evaluate expectation values of various operators in this state:
\begin{equation}
\langle {\mathcal O}\rangle=\int D\phi \Psi^*[\phi]{\mathcal O}\Psi[\phi]\, .
\end{equation}
A calculation of this kind is, obviously, tantamount to the
evaluation of a Euclidean functional integral with the square of
the wave functional (WF) playing the role of the partition
function. One should therefore be able to solve exactly a
$d$-dimensional field theory with the action
\begin{equation}
S[\phi]=-{\rm log}\Psi^*[\phi]\Psi[\phi]\, .
\end{equation}
In quantum mechanics such a concern plays only a background role.
Whatever the chosen trial state might be, the calculation to be
performed involves integrals of functions. The evaluation of any
such integral can be tackled, if not always analytically then
numerically, without major complications. In QFT, where our
ability to evaluate path integrals is, to say the least, limited,
the calculability restriction on the trial wave functional is very
severe. Since in dimension $d>1$ the only theories one can solve
exactly are free field theories, the requirement of calculability
almost unavoidably restricts the possible form of the trial WF to
a Gaussian state:
\begin{equation}
\Psi[\phi]=\exp\left\{-\frac{1}{2}
\int d^{3}x d^{3}y \left[ \phi(x)-\zeta(x)\right]
G^{-1}(x,y)\left[\phi(y)-\zeta(y)\right]\right\}
\label{gaus}
\end{equation}
with $\zeta(x)$ and $G(x,y)$  being c-number functions.

Another serious problem is that of ``ultraviolet modes". The main
motivation of a variational calculation in a strongly interacting
theory is to learn about the distribution of the low momentum
modes of the field in the vacuum wave functional. However, the VEV
of the energy (and all other intensive quantities) is dominated
entirely by contributions of high momentum fluctuations, for the
simple reason that there are infinitely more ultraviolet modes
than modes with low momentum. Therefore, even if one has a very
good idea of how the WF at low momenta should look like, if the
ultraviolet part of the trial state is even slightly incorrect the
minimization of energy may lead to absurd results. Due to the
interaction between the high and low momentum modes, there is a
good chance that the infrared (IR) variational parameters will be
driven to values which minimize the interaction energy, and have
nothing to do with the dynamics of the low momentum modes
themselves.

Finally, in gauge theories there is an additional complication: that  of 
 gauge invariance. The allowed wave functions must be invariant
under the time independent gauge transformations. If one does not
impose the Gauss' law on the states exactly, one is not solving
the right problem. The QCD Hamiltonian is only defined on the
gauge invariant states, and its action on non gauge invariant
states can be modified at will. Thus, by minimizing a particularly
chosen Hamiltonian without properly restricting the set of allowed
states, one is taking the risk of  finding a  ``vacuum'' which has
nothing to do with the physical one, but is only picked due to a
specific form of the action of the Hamiltonian outside the
physical subspace. There are two ways to approach this problem.
One is to solve Gauss' law operatorially {\it \`a la} Dirac
\cite{Dirac:pj,Dirac:1958jc}. This leads to a ``completely gauge fixed''
formalism. Once this is done any state can be chosen. The price to
be paid is that the Hamiltonian in any completely fixed gauge is
very complicated. The calculation of the expectation value of
energy then is not analytically manageable. Another way is to let
the Gauss' law constraint be, but write down trial states which
are explicitly gauge invariant. The Hamiltonian in such a
calculation is simple, but the trial states are usually very
complicated. Thus the problem of gauge invariance is linked very
strongly with the problem of calculability.

\section{The variational setup}

So given that the only path integral that we know how to calculate
analytically is that of a Gaussian wave function, what kind of
progress can one make with variational calculations in QFT?
Surprisingly enough in QFT without gauge symmetry a Gaussian
ansatz can take one a long way. The famous BCS theory of
superconductivity is nothing but a variational analysis of the
interacting QFT using a Gaussian variational state\cite{bcs}. The
Gaussian Ansatz has also been applied to self-interacting
relativistic scalar and spinor theories \cite{moshe}, where it
gives non-trivial exact results in the large $N$ limit. The reason
it works is quite simple. A Gaussian wave functional is the exact
ground state in quantum field theories of non-interacting fields,
massless or massive. As a trial state it is thus flexible enough
to probe the existence of a mass gap in the spectrum. Therefore,
whenever the main effect of the interaction is to dynamically
generate (or significantly change) the mass of the particles, the
Gaussian Ansatz is adequate and informative. Put another way, one
can hope that the Gaussian WF is useful if the non-perturbative
physics is dominated by a single condensate, that of the lowest
possible dimension.
 From this point of view, it would seem that, it is perfectly
reasonable to try a similar variational Ansatz in the Yang-Mills
theory. After all, it is strongly suggested by the QCD sum rules
\cite{Shifman:1979bx} that the pure glue sector is strongly
dominated by the lowest dimensional non-perturbative condensate
$\langle F^2\rangle$.

The difficulty  comes from the necessity to impose gauge invariance.
 It is very
easy to see that in a non-abelian theory it is impossible to write
a Gaussian WF that satisfies the constraint of gauge invariance.
The SU(N) gauge theory is described by a Hamiltonian
\begin{equation}
{\mathsf H}= \int d^{3}x \left[\frac{1}{2}E^{a2}_i+\frac{1}{2}B^{a2}_i\right]
\end{equation}
where
\begin{align}
E^a_i(x)&=i\frac{\delta}{\delta A^a_i(x)} \, ,\nonumber \\
B^a_i(x)&=\frac{1}{2}\epsilon_{ijk}
\{\partial_jA_k^a(x)-\partial_kA^a_j(x)+gf^{abc}A_j^b(x)A_k^c(x)\}\, ,
\end{align}
and all physical states must satisfy the constraint of gauge
invariance
\begin{equation}
G^a(x)\Psi[A]=\left[\partial_iE^a_i(x)-gf^{abc}A^b_i(x)E^c_i(x)\right]
\Psi[A]=0\, .
\label{constr}
\end{equation}
Under a gauge transformation $U$ --- generated by $G^a(x)$ --- the
vector potential transforms as
\begin{equation}
A^a_i(x)\rightarrow   A^{Ua}_i(x)=S^{ab}(x)A_i^b(x)+\lambda_i^a(x)
\label{gt}
\end{equation}
where
\begin{eqnarray}
S^{ab}(x)=\frac{1}{2}\tr\left(\tau^aU^\dagger\tau^bU\right);
\ \ \ \lambda_i^a(x)=\frac{i}{g}\tr\left(\tau^aU^\dagger
\partial_iU\right)
\end{eqnarray}
and $\tau^a$ are traceless Hermitian N by N matrices satisfying
$\tr(\tau^a\tau^b)=2\delta^{ab}$. A  Gaussian wave functional
\begin{multline}
\Psi[A_i^a]=\exp\Big\{ - \frac{1}{2}\int d^3 x d^3 y
 \left[A_i^a(x)-\zeta_i^a(x)\right]\\
\cdot (G^{-1})^{ab}_{ij}(x,y)
 \left[A_j^b(y)-\zeta_j^b(y)\right]\Big\}
 \label{eq:psiA}
\end{multline}
transforms under the gauge transformation as
\begin{equation}
\Psi[A_i^a]\rightarrow\Psi[(A^{U})_i^{a}] \, .
\end{equation}
In the abelian case it is enough to take $\partial_iG^{-1}_{ij}=0$
to satisfy the constraint of gauge invariance.  In the non-abelian
case, however, due to the homogeneous piece in the gauge transformation
eq.~(\ref{gt}), no gauge invariant Gaussian WF exists.
Thus one has to abandon the notion of a simple Gaussian variational Ansatz on the Hilbert space
spanned by the canonical variables $\{A^a_i(x),E^a_i(x)\}$.

One option is to attempt to  solve the constraint
eq.~(\ref{constr}) operatorially. Such a procedure leads to a
``completely gauge fixed'' formalism. For example if one solves
eq.~(\ref{constr}) for $E_3^a$ and then performs Dirac
quantization, one obtains the Hamiltonian in the axial gauge.
Another popular choice is to express $\partial_iE_i^a$ in terms of
the transverse components of electric field and vector potential.
This leads to the Hamiltonian in the Coulomb gauge. In either
case, after the Hilbert space has been reduced to the physical
space, the Gaussian trial state in the remaining degrees of
freedom  can be considered. However, the actual calculations are
extremely difficult since the Hamiltonian in any completely fixed
gauge is very complicated. Attempts to perform such a calculation
in the axial gauge were made in
\cite{Nojiri:1984nu,Rosenstein:1986ku}. The problem 
encountered here is that the calculation is plagued by a spurious
infrared pole of the form $1/k_3$. The results depend very much on
the way one treats this pole, and there is no clear understanding
as to the proper way to do so. The Coulomb gauge calculation is
even less straightforward, primarily because the Coulomb gauge is
known to suffer from Gribov ambiguity \cite{Gribov:1978wm}. Thus
the expression for the Hamiltonian that one obtains in this gauge
is only formal, and has to be very carefully defined by
considering zeros of the Fadeev-Popov operator. Non-perturbative
analytic calculations in this gauge are therefore all but
impossible \cite{Schutte:1985sd}. Attempts have been made to
perform numerical variational minimization taking account of the 
Gribov horizon \cite{Cutkosky:1987nh,Cutkosky:1988yi}, but the
interpretation of the results is very difficult.

Another route, pursued in \cite{Greensite:1979yn}, is to modify the Gaussian trial state in a ``minimal'' way to
make it compatible with gauge invariance. Thus one considers the state
\begin{equation}
\Psi[A]=\exp\left\{ -\int d^3 x d^3 y\, B^a_i(x)W^{ab}(x,y)B^b_j(y)G_{ij}(x-y)\right\}
\end{equation}
where $W^{ab}$ is the Wilson path ordered integral along the straight line connecting points $x$ and $y$
taken in the adjoint representation. Explicit introduction of the Wilson line indeed makes the state gauge invariant.
Some semi-quantitative arguments were given in \cite{Greensite:1979yn} to the effect that the best variational state of this
type will have
a short range variational function $G(x-y)$ and therefore will be confining at large distances.
However the state is so complicated that
no reasonable way to get the actual calculation going was ever found.

Finally, another series of works attempts to initially disregard the Gauss' law constraint, and subsequently  
calculate corrections due to its implementation perturbatively \cite{Consoli:1985sx,Preparata:1986bk,Preparata:1986cq,Kerman:1989kb,Heinemann:1999ja}.
A clear discussion of this method and the analogy with nuclear physics calculations
is given in a recent review paper \cite{Schroeder:2003qw}.
Although the method is mathematically very elegant, it is not suited for systems
whose energy may be lowered by a non-perturbative amount due to the exact implementation of  Gauss' law.
As we believe this to be the case for the Yang-Mills theory, we strongly doubt the usefulness of this method
to obtain realistic results.

In this review we will therefore restrict ourselves to the
discussion of the approach proposed in \cite{Kogan:1995wf}. The
scheme devised in \cite{Kogan:1995wf} is rather straightforward.
One opts for strictly preserving gauge invariance, and constructs
a gauge invariant wave functional by projecting the Gaussian wave
functional of eq.~(\ref{eq:psiA}) onto the gauge invariant sector.
Restricting ourselves to the case of zero classical fields
($\zeta=0$), the variational Ansatz proposed in
\cite{Kogan:1995wf} is
\begin{equation}
\Psi[A_i^a]=\int DU(x)
 \exp\left\{-\frac{1}{2}\int d^3 x d^3 y
\, A_i^{Ua}(x)G^{-1ab}_{ij}(x,y)\ A_j^{Ub}(y)\right\}
\end{equation}
with $A_i^{Ua}$ defined in eq.~(\ref{gt}) and the integration
is performed over the space of special unitary matrices with
the $SU(N)$ group invariant measure.

The trial state defined in this way is explicitly gauge invariant. It is not obvious at this stage that we will be
able to deal with it analytically. But as we will see in the following, some headway can be made using analytical
approximation schemes to calculate various expectation values in this state.
In this review we will discuss the applications of this technique to the pure Yang-Mills theory at zero as well as at finite
temperature.

But before plunging head first into the discussion of non-abelian Yang-Mills theories, we will consider the much
simpler, but nevertheless informative, case of a compact U(1) gauge theory in 2+1 dimensions \cite{Kogan:1995vb}.
The theory is known to be
confining, and our aim is to see whether the confining properties of the ground state can be captured with our
simple Ansatz.
It will also illustrate how the use of different field theoretical techniques
allows one to carry out the variational estimation of the ground state.

\section{Compact QED in 2+1 dimensions}
\label{sec:qed}

We start by setting up the Hamiltonian description of U$(1)$ compact QED.
First of all, we need to determine what the  Hilbert space of admissible states is.
It is clear that Gauss' law should be implemented, and thus all the physical states should satisfy
\begin{equation}
\exp\left\{i\int
d^2x\partial_i\phi(x)E_i(x)\right\}|\Psi\rangle=|\Psi\rangle
\label{constrqed}
\end{equation}
for an arbitrary continuous function $\phi(x)$. This is true for
both compact and non-compact QED. There is however a crucial
difference between Gauss' law in the compact theory and in the
non-compact one. In the non-compact theory eq.~(\ref{constrqed})
should be satisfied only for regular functions $\phi$. For
example, the operator
\begin{equation}
V(x)=\exp\left\{\frac{i}{g}\int d^2y \, \epsilon_{ij}
\frac{(x-y)_j}{(x-y)^2} E_i(y)\right\}
\label{vortex}
\end{equation}
which has the form of  eq.~(\ref{constrqed}) with the function
$\phi$ proportional to the planar angle $\theta$, i.e.
$\phi=\frac{1}{g} \theta(x)$, does not act trivially on the
physical states. In fact, this operator creates a point-like
magnetic vortex with magnetic flux $2\pi/g$ and therefore changes
the physical state on which it acts.

In the compact theory the situation in this respect is quite different.
Point-like vortices with quantized magnetic flux $2\pi n/g$ cannot be detected by any measurement.
This translates into the requirement that the creation operator of a point-like vortex must be indistinguishable from the unit operator.
In other words, the operator eq.~(\ref{vortex}) generates a transformation which belongs to the
compact gauge group, and should therefore act trivially on all physical states.
Eq.~(\ref{constrqed}) should therefore be satisfied also for these operators.

Accordingly, the Hamiltonian of the compact theory must be also invariant under these transformations.
The magnetic field itself, defined as $B=\epsilon_{ij}\partial_iA_j$, does  not commute
with $V(x)$ (cf. ref. \cite{vort}),
\begin{equation}
V^\dagger (x)B(y)V(x)=B(y)+\frac{2\pi}{g}\delta^2(x-y)\, .
\label{commut}
\end{equation}
The Hamiltonian should therefore contain not $B^2$ but rather a periodic function of $B$.
We will choose our Hamiltonian to be
\begin{equation}
H=\frac{1}{2}a^2\sum E^2_{\bn i}-\frac1{g^2a^2}\sum \cos
ga^2B_{\bn}\, .
\label{ham}
\end{equation}
Since we will need in the following an explicit ultraviolet
regulator, we switched temporarily to lattice notations. Here $a$
is the lattice spacing, and the sums are respectively over the
links and plaquettes of the two-dimensional spatial lattice. The
coefficients of the two terms in the Hamiltonian are chosen so
that in the weak coupling limit, upon formal expansion to lowest
order in $g^2$, the Hamiltonian reduces to the standard free
Hamiltonian of 2+1 dimensional electrodynamics. Following Polyakov
\cite{Polyakov:1987ez}, we work in the weakly coupled regime.
Since the coupling constant $g^2$ in 2+1 dimensions has dimension
of mass, weak coupling means that the following dimensionless
ratio is small
\begin{equation}
g^2a\ll 1.
\end{equation}

Our aim now is to find variationally the vacuum wave functional of this theory.

\subsection{The variational Ansatz}

As our variational trial ground state we choose the Gaussian wave function of $A$ projected onto the gauge singlet.
The projection has to be performed with respect to the full compact gauge group of eq.~(\ref{constrqed}).

To facilitate this, we define a vortex field $A^V(x)$ that satisfies (we suppress the lattice spacing $a$ henceforth)
\begin{equation}
\nabla\times A^V(x)=\frac{2\pi}{g}\delta^2(x)
\ ,\qquad \nabla\cdot A^V=0\, .
\label{vortex1}
\end{equation}
This is the vector potential corresponding to a magnetic field that is zero everywhere except at $x=0$, where it takes the value $\frac{2\pi}g$.
The explicit solution of eq.~(\ref{vortex1}) is
\begin{equation}
A^V_i(x)=-\frac{1}{g}\epsilon_{ij}\frac{x_j}{x^2}\, .
\end{equation}
The compact gauge invariance requires the variational wave function $\psi[A]$ to be invariant under shifts $A\to A+A^V$.
This is, of course, consistent with the periodicity of $H$ under $B\to B+\frac{2\pi}g$.

Hence we define a field, shifted by a non-compact gauge transformation $\phi_\bn$ and by an integer valued  vortex distribution $m(x)$,
\begin{equation}
A^{(\phi,m)}(x)=A(x)-\nabla\phi(x)
-\sum_{y}m(y)A^V(x-y)
\label{shift1}
\end{equation}
or, for short,
\begin{equation}
A^{(\phi,m)}=A-\nabla\phi-A^V\cdot m\, .
\label{shift2}
\end{equation}
We choose the gauge invariant and periodic trial wave function as
\begin{equation}
\psi[A]=\sum_{\{m(x)\}}\int[d\phi]\,
\exp\left[-\frac{1}{2}\int d^2x d^2yA^{(\phi,m)}_{x}
G^{-1}(x-y)A^{(\phi,m)}_{y}\right]\, .
\label{wf}
\end{equation}
Under a gauge transformation,
\begin{equation}
\psi[A+\nabla\lambda]=\psi[A] \label{g_invariance}
\end{equation}
since $\lambda$ can be absorbed into a shift in $\phi$.
The simple rotational structure of $G_{ij}=\delta_{ij}G$ that appears in the variational wave function eq.~(\ref{wf}) is consistent with perturbation theory.
We also take $G(x)$ to be a real function.

Our task now is to calculate the expectation value of the Hamiltonian in this state, and to minimize it with respect to the variational function $G$.
We start by considering the normalization of the wave function.

\subsection{The normalization integral}

The norm of $|\psi\rangle$ is
\begin{multline}
Z  \equiv \langle\psi|\psi\rangle
= \sum_{\{m,m'\}}\int[d\phi][d\phi'][d\bA]\\
\times
\exp\left[{-\frac12A^{(\phi,m)}G^{-1}A^{(\phi,m)}}\right]
\exp\left[{-\frac12A^{(\phi',m')}G^{-1}A^{(\phi',m')}}\right]\, .
\label{norm1}
\end{multline}
We shift $A$ by $\nabla\phi'+A^V\cdot m'$ and absorb the shift into $\phi$ and $m$, obtaining
\begin{equation}
Z=\sum_{\{m\}}\int[d\phi][d\bA]\,
e^{-\frac12A^{(\phi,m)}G^{-1}A^{(\phi,m)}} e^{-\frac12AG^{-1}A}\, .
\label{Z}
\end{equation}
Now we combine the exponents according to
\begin{multline}
A^{(\phi,m)}G^{-1}A^{(\phi,m)}+AG^{-1}A = 2A^{(\phi/2,
m/2)}G^{-1}A^{(\phi/2, m/2)} \\
+\frac12 S(\phi,m)G^{-
1}S(\phi,m)\, ,
\end{multline}
where
\begin{equation}
S\equiv \nabla\phi+ A^V\cdot m\, .
\end{equation}
It is easy to see that $SG^{-1}S$ contains no cross terms between $m$ and $\phi$.
We shift $A$ by $\nabla\phi/2+A^V\cdot m/2$, and all the fields decouple.
We have then
\begin{equation}
Z=Z_A Z_{\phi} Z_v\, ,
\end{equation}
where
\begin{align}
Z_A&=\det \pi G\, ,\\
Z_{\phi} &=\int[d\phi]\, e^{-\frac14\nabla\phi\cdot
G^{-1}\cdot\nabla\phi} =\left(\det
4\pi\frac1{\nabla^2}G\right)^{1/2}\, ,\\
Z_v&=\int_{\{m_{\bn'}\}}\exp\left[-
\frac1{4g^2}\int d^2xd^2y\, m(x)D(x-y)m(y) \right]\, . \label{Zv1}
\end{align}
Here, $Z_v$ is the vortex ``partition function", with the
``vortex-vortex interaction'' $D$ given by
\begin{equation}
D(x'-y')=g^2\int d^2xd^2y \, A^V(x-x')\cdot
G^{-1}(x-y) A^V(y-y')\, .
\end{equation}
We can split off the $x=y$ terms in eq.~(\ref{Zv1}) and write
\begin{equation}
Z_v=\int [dm(x)]\exp\left[-\frac1{4g^2}\int_{x\not=y}m(x)
D(x-y)m(y) \right] \prod_{y}z^{m(y)^2}
\label{Zv2}
\end{equation}
where we have defined the vortex fugacity
\begin{equation}
z=e^{-\frac1{4g^2}D(0)}\, .
\label{fugacity1}
\end{equation}

We expect the ultraviolet behaviour of the variational function $G$
to be the same as in the free theory, {\em viz.} (for the Fourier
transform),
\begin{equation}
G^{-1}(k)\sim k\, ,
\end{equation}
so that
\begin{equation}
D(0)\sim\int^\Lambda\frac{d^2k}{(2\pi)^2}\frac{4\pi^2}{k^2}G^{-1}(k)
\label{DG}
\sim 2\pi\Lambda
\end{equation}
and thus
\begin{equation}
z\sim e^{-\frac{\pi}2\frac{\Lambda}{g^2}}\, ,
\label{potential}
\end{equation}
where we use the momentum space cutoff $\Lambda=a^{-1}$.

In the weak coupling region we have $z\ll1$, which allows us to
restrict possible values of $m$ to $0,\pm1$ in
eqs.~(\ref{Zv1},\ref{Zv2}).

\subsection{Expectation values}

We are now ready to calculate the expectation value of the Hamiltonian eq.~(\ref{ham}).
Using the definition eq.~(\ref{wf}), we obtain
\begin{align}
V^{-1}\eval{\int E^2\,d^2x}
&=-\frac1V\eval{\psi\left|\sum_{\bn,i}
\frac{\partial^2}{\partial  A_{\bn,i}^2}\right|\psi}\nonumber\\
&=\frac12\int \frac{d^2k}{(2\pi)^{2}} G^{-1}(k)-
\frac{\pi^2}{g^2}
\int \frac{d^2k}{(2\pi)^{2}}  k^{-2}G^{-
2}(k)K(k)\label{elen1}\, ,
\end{align}
where $K(k)$ is the correlation function of the vorticity
\begin{equation}
K(k)\equiv\int d^2x\,e^{ikx}\eval{m(x)m(0)}\, .
\end{equation}

To calculate correlation functions of $m$ we use a duality transformation \cite{Coleman:1975bu,Samuel:1978vy,Amit:1980ab,Boyanovsky:1989ge}.
 We add an $iJ\cdot m$ term to the exponent in eq.~(\ref{Zv1}) and use the formula
\begin{equation}
e^{-\frac1{4g^2} m\cdot D\cdot m} ={\it const}\int[d\chi]\,
e^{-g^2\chi\cdot D^{-1}\cdot\chi} e^{i\chi\cdot m}
\label{duality}
\end{equation}
to obtain
\begin{equation}
Z_v[J]=\int[d\chi]\, e^{-g^2\chi\cdot D^{-1}\cdot\chi}
\prod_{x}
[1+2\cos(\chi(x)+J(x))]\, .
\label{sin1}
\end{equation}
Noting that\,\footnote{The normal ordering is performed relative to the free theory defined by the quadratic action in eq.~(\ref{sin1}).}
\begin{equation}
\cos (\chi+J)=\eval{\cos\chi}_0:\!\cos
(\chi+J)\!:\,=z:\!\cos
(\chi+J)\!: \label{normalorder}
\end{equation}
we have
\begin{align}
Z_v&=\int[d\chi]\, e^{-g^2\chi\cdot D^{-1}\cdot\chi}
\prod \left[1+2z:\!\cos (\chi+J)\!:\right]\nonumber\\
&\simeq \int D\chi\exp\left[-g^2\chi D^{-1} \chi+ 2z\int
d^2x\,:\!\cos\Big(\chi(x)+J(x)\Big)\!:\right]\, .
\label{sine}
\end{align}
Correspondingly \cite{Kogan:1995vb},
\begin{equation}
\eval{m(x)m(y)}=2g^2D^{-1}(x-y)-4g^4\eval{D^{-1}\chi(x)\,D^{-1}\chi(y)}
\end{equation}
and
\begin{equation}
K(k)=2z + O(z^2)\, ,
\label{rhoc}
\end{equation}
which, in this approximation, does not depend on momentum.

The propagator of $\chi$ is also easily calculated.
To first order in $z$,
\begin{align}
\int d^2x\,e^{ikx}\eval{\chi(x)\chi(0)}
&= \frac{1}{2g^2D^{-1}(k)+2z}\nonumber\\
&=\frac{D(k)}{2g^2} -z\frac{D^2(k)}{2g^4} + O(z^2)\, .
\end{align}

Thus we have
\begin{equation}
V^{-1}\eval{\int E^2\,d^2x}=\frac{1}{2}\int \frac{d^2k}{(2\pi)^{2}}
G^{-1}(k)-\frac{2\pi^2}{g^2} z\int \frac{d^2k}{(2\pi)^{2}}
k^{-2}G^{-2}(k)\, .
\label{elen}
\end{equation}

The magnetic part is easily calculated since it has an exponential form and, therefore, with our trial wave function leads to a simple Gaussian integral.
We find
\begin{equation}
\eval{e^{ingB_{\bn}}}=\exp\left[-\frac{1}{4}n^2g^2
\int\frac{d^2k}{(2\pi)^2}k^2G(k)\right]\eval{e^{in\pi m_{\bn}}}\, .
\end{equation}
The second factor is different from unity only for odd values of
$n$. Using eq.~(\ref{sine}) we find $\eval{e^{i\pi m}}=e^{-4z}$.
Expanding to leading order in $g^2$ and $z$, we get
\begin{equation}
\eval{-\frac{1}{g^2}\cos gB} = \frac{1}{4}\int \frac{d^2k}{(2\pi)^2}
k^2G(k)+\frac{4}{g^2}z\, ,
\label{Benergy}
\end{equation}
where we have dropped an additive constant. Finally, the
expression for the variational vacuum expectation value of the
energy to first order in $z$ is
\begin{equation}
\frac{1}{V}\eval H= \frac{1}{4} \int\frac{d^2k}{(2\pi)^2}
\left[G^{-1}(k)+k^2G(k) -\frac{4\pi^2}{g^2}
z\left(k^{-2}G^{-2}(k)- \frac{4}{\pi^2}\right) \right]\, .
\label{eval}
\end{equation}

\subsection{Determination of the ground state}

The expression eq.~(\ref{eval}) has to be minimized functionally
with respect to $G(k)$. From eqs.~(\ref{fugacity1},\ref{DG})  we
find
\begin{equation}
\frac{\delta z}{\delta G(k)}=\frac{1}{4g^2}k^{-2}G^{-2}(k)\,z\, .
\end{equation}
The variation of eq.~(\ref{eval}) gives\,\footnote{We have dropped  a term $-\frac{8\pi^2}{g^2}zk^{-2}G^{-3}(k)$ from the right-hand side of eq.~(\ref{variation}) since it  is smaller by a factor of $\frac{g^2k}{\Lambda^2}$ than the term retained when one assumes $G\sim k^{-1}$ at large $k$.}
\begin{equation}
{k^2-G^{-2}(k) = \frac{4\pi^4}{g^{4}}
z k^{-2} G^{-2}(k) \int \frac{d^2p}{(2\pi)^2}
\left[p^{-2}G^{-2}(p)- \frac{4}{\pi^2} \right]}\ .
\label{variation}
\end{equation}
Eq.~(\ref{variation}) has the solution
\begin{equation}
G^{-2}(k)=\frac{k^4}{k^2+m^2}\, ,
\label{solution}
\end{equation}
where
\begin{equation}
{m^2= \frac{4\pi^4}{g^4}
z\int \frac{d^2k}{(2\pi)^2} \left[k^{-2}G^{-2}(k)-
\frac{4}{\pi^2}\right]} \, .
\label{mass}
\end{equation}
The main contribution to the integral in the gap equation, eq.~(\ref{mass}), comes from momenta $k^2\gg m^2$.
For these momenta $k^2G^{-2}(k)=1$.
Therefore, we see that eq.~(\ref{mass}) has a non-trivial solution.
Using eqs. (\ref{fugacity1},\ref{DG},\ref{solution}) we obtain
\begin{equation}
{m^2= \frac{4\pi^4}{g^4}
\exp\left(-\frac{\pi^{2}}{g^{2}} \int \frac{d^2p}{(2\pi)^2}
\frac{1}{\sqrt{p^2 + m^{2}}}\right) \int
\frac{d^2k}{(2\pi)^2}
\left[\frac{k^2}{k^2+m^2}-\frac{4}{\pi^2}\right]}
\label{complicatedm}
\end{equation}
which for $g^2 a = g^2/\Lambda \ll1$ can be simplified to (cf. eq~(\ref{potential}))
\begin{equation}
{m^2 = {4\pi^2} \frac{(\pi^2-4)\Lambda^4}{g^4}
\exp\left(-\frac{\pi\Lambda }{2g^2}\right)}\, ,
\label{hamiltonianmassm}
\end{equation}
where we have restored the ultraviolet cutoff dependence explicitly.
The resulting $m$ is the mass gap of the theory, in the sense that it is the inverse of the spatial correlation length.
Calculating, for example, the propagator of magnetic field, we find
\begin{equation}
\eval{e^{igB_{\bm}}e^{-
igB_{\bn}}}=\left|\eval{e^{igB}}\right|^2
e^{\frac{g^2}2\nabla^2G(\bm-\bn)}\, ,
\end{equation}
and at large distances (neglecting power-like prefactors),
\begin{equation}
\nabla^2G(x)=-
\int\frac{d^2k}{(2\pi)^2}(k^2+m^2)^{1/2}e^{i\bk\cdot\bx}
\sim e^{-mx} \, .
\end{equation}
This dynamically generated mass is Polyakov's result \cite{Polyakov:1975rs,Polyakov:1977fu}.
Thus, we recover in the Hamiltonian approach the first important result known about compact QED ---  a finite mass gap $m$, as well as its correct dependence on the coupling constant.

\subsection{Spatial Wilson loops}

We also want to see whether the charges are confined in our best variational state.
The simplest quantity that is related to confinement is the expectation value of the Wilson loop.
Therefore, we will calculate it in our ground state
\begin{equation}
W_l[C]= \eval{\exp\left(ilg\oint_C \bA\cdot d\bx\right)} =
\eval{\exp\left(ilg\int_\Sigma B\,dS\right)}\, ,
\end{equation}
where $l$ is an integer and the integral is over the area $\Sigma$ bounded by the loop $C$.
We have
\begin{equation}
W_l[C]=\eval{\prod_S e^{il\pi m_\bn}}Z_A^{-1}\int D\bA
\exp\left(
-AG^{-1}A+ilg\int_\Sigma B\,dS\right)\, .
\label{wilson}
\end{equation}
The second factor is a Gaussian integral, which gives
\begin{equation}
W_A=\exp\left(\frac{l^2g^2}4\int_\Sigma d^2x\int_\Sigma
d^2y\,
\nabla^2G(\bx-\by)\right)\, .
\label{wa}
\end{equation}
In the limit of large $\Sigma$ the leading behaviour of the exponent is
\begin{equation}
-\frac{l^2}4g^2\Sigma\lim_{k\to0}k^2G(k)=-
\frac{l^2}{4}g^2m\Sigma\, .
\end{equation}
This gives the area law with the string tension
\begin{equation}
\sigma=\frac{l^2}{4}g^2m\, .
\label{str}
\end{equation}
The first factor in eq.~(\ref{wilson}) is different from unity only for odd $l$.
It can be easily calculated but gives only subleading corrections to the string tension \cite{Kogan:1995vb}.

\subsection{Potential between external charges}

We have thus found that in the best variational state the Wilson
loop has an area law behaviour. This usually signals confinement,
and one is naturally inclined to conclude that we have indeed
found confinement with the string tension related in the expected
way to the dynamically generated scale, $\sigma\propto g^2m$. One
must be however a tad more careful at this stage, since the
spatial Wilson loop does not directly give the potential between
external charges. Although in the Euclidean formulation there is
no difference between spatial and time-like Wilson loops, in the
Hamiltonian approach this is not obvious. It is, therefore,
desirable to calculate directly the potential between external
charges.

How does one do this? Obviously one has to introduce into the
theory the source corresponding to the pair of external charges.
The result is a modification of Gauss' law, \be
\partial_iE_i(x)=g\rho(x)
\ee with $\rho(x)=\delta(x-x_1)-\delta(x-x_2)$. As the external
charges do not have dynamics of their own, the Hamiltonian remains
unchanged. The seemingly simplest option seems to be to take the
same Gaussian variational state which minimizes the energy in the
vacuum sector, and project it with the modified projection
operator corresponding to the new Gauss' law. This ``minimally
modified'' state would be
\begin{multline}
\psi[A]=\sum_{\{m(x)\}}\int[d\phi]
\exp \bigg[-\frac12\int d^2x d^2yA^{(\phi,m)}_{x}
G^{-1}(x-y)A^{(\phi,m)}_{y}\\
+ig\rho(x)\phi(x)\bigg]
\, . \label{wfmm}
\end{multline}
One could then calculate the energy expectation value in this state and take this as an estimate of the
interaction potential. This procedure was suggested in \cite{Diakonov:1998ir,Zarembo:1998ms,Zarembo:1998xq}.
However it turns out that the estimate obtained for the interaction energy in this way is very
unreliable. For example, in the compact QED$_3$ case this calculation was performed in \cite{Kovner:1998eg} with the resulting energy being not even infrared finite.
Instead, to get a reasonable estimate for the energy one has to introduce additional variational parameters.
In \cite{Kovner:1998eg} the variational Ansatz was extended in the following way:
\begin{multline}
\psi[A]=\sum_{\{m(x)\}}\int[d\phi]
\exp \bigg[-\frac12\int d^2x d^2yA^{(\phi,m)}_{x}
G^{-1}(x-y)A^{(\phi,m)}_{y}\\
+ie_i(x)A^{(\phi,m)}_i+ig\rho(x)\phi(x)\bigg] \, .
\label{wfc}
\end{multline}
Here, the ``classical'' electric field profile $e_i(x)$ is to be varied so as to minimize the energy.
It turns out that with this extra variational parameter the calculation gives  very satisfactory results \cite{Kovner:1998eg}.
Without giving the details of this calculation, we note that it confirms the expectation that the potential between external charges is a linear function of the separation.
The  string tension calculated this way coincides within ten percent with eq.~(\ref{str}).

What have we learned from this toy model?
First of all it is very encouraging that the Gaussian projected Ansatz is good enough to reproduce all known results and, in particular, confinement and dynamical mass generation with parameterically correct values of string tension and mass.

We have also seen that we need some ingenuity to be able to carry out the calculations with the variational wave function.
We were able to do it in this simple case as the coupling constant was small
 and we could utilize existing methods for treating weakly interacting two dimensional systems.
It is quite clear that in 3+1 dimensions no such techniques will
be available and consequently the situation will be, in this
respect, much more complicated. Finally, we learned that we could
not impose the Gauss' law constraint on the state perturbatively.
Perturbatively, the operator $V$ --- eq.~(\ref{vortex}) --- simply
vanishes. However, without including it into the projection
procedure of the Gaussian Ansatz, we would not get any non-trivial
results --- we would have found the vacuum of a free massless
photon without a dynamical mass gap and with vanishing string
tension. We fully expect that this aspect will remain important in
the application to QCD, to which we turn in the next section.

\section{The Yang-Mills theory}
\label{sec:ym}

The dynamics of the pure glue sector of QCD is described by the Yang-Mills (also often called gluodynamics) Hamiltonian
\begin{equation}
{\mathsf H}= \int d^{3}x \left[\frac{1}{2}E^{a2}_i+\frac{1}{2}B^{a2}_i\right]
\label{ymham}
\end{equation}
where
\begin{align}
E^a_i(x)&= i \frac{\delta}{\delta A^a_i(x)}\, , \nonumber \\
B^a_i(x)&= \frac{1}{2}\epsilon_{ijk}\,
\{\partial_j A_k^a(x)-\partial_kA^a_j(x)+gf^{abc}A_j^b(x)A_k^c(x)\}
\end{align}
and all physical states must satisfy the constraint of gauge invariance
\begin{equation}
G^a(x)\Psi[A]=\left[\partial_iE^a_i(x)-gf^{abc}A^b_i(x)E^c_i(x)\right]
\Psi[A]=0 \, .
\label{gaussconstr}
\end{equation}

We thus have to choose a set of variational states which are invariant under the action of eq.~(\ref{gaussconstr}).
We start then with a Gaussian state
\begin{equation}
\Psi_0[A_i^a]=\exp\left\{ - \frac{1}{2}\int d^{3}x d^{3}y\, 
 A_i^a(x)
 (G^{-1})^{ab}_{ij}(x,y)
 A_j^b(y)\right\}\, ,
 \label{psi0}
\end{equation}
where the set of functions  $G^{ij}_{ab}(x)$ are variational parameters.

We hope that the freedom allowed by variation of $G^{-1}$ is sufficiently wide to probe the
non-perturbative physics of the Yang-Mills vacuum. As discussed in the introduction, the states of the form eq.~(\ref{psi0}) are not gauge invariant and, therefore, as such do not belong to the physical Hilbert space of gluodynamics.
To remedy this problem we project the states onto the gauge invariant subspace by gauge transforming them and integrating over the whole gauge group\,\footnote{Hereafter we use the notational shorthand $\int_x \equiv \int d^3 x$}:
\begin{equation}\Psi[A_i^a]=\int DU(x)
 \exp\left\{-\frac{1}{2}\int_{x,y}
\ A_i^{Ua}(x)G^{-1ab}_{ij}(x,y)\ A_j^{Ub}(y)\right\}
\label{ansatz}
\end{equation}
with $A_i^{Ua}$ defined as
\begin{equation} \ A^{Ua}_i(x)=S^{ab}(x)A_i^b(x)+
\lambda_i^a(x)\, ,
\label{gtqcd}
\end{equation}
and
\begin{equation}
S^{ab}(x)=\frac{1}{2}\tr\left(\tau^aU^\dagger\tau^bU\right);
\ \ \ \lambda_i^a(x)=\frac{i}{g}\tr\left(\tau^aU^\dagger
\partial_iU\right)
\label{defin}
\end{equation}
with $\tau$ --- traceless $ N\times N$ hermitian matrices --- normalized by
\begin{equation}
    \tr (\tau^a \tau^b) = 2 \delta^{ab}\, ,
\end{equation}

The integration in eq.~(\ref{ansatz}) is performed over the space of special unitary matrices with the $SU(N)$ group invariant measure.
This integration projects the original Gaussian state onto a colour singlet.
Due to the projection operation, the calculation of expectation values in this state is much more involved than in the case of a simple Gaussian.
A full functional minimization with respect to the variational functions $G^{-1ab}_{ij}(x,y)$ is, therefore, beyond our calculational abilities.

In order to render calculations possible, we will impose several restrictions on the form of $G^{-1}$, which will lead to considerable simplifications.
First, we require the state to be translationally invariant, that is we assume that Lorentz symmetry is not spontaneously broken in the ground state, restricting the form of $G^{-1}$
\begin{equation}
G^{-1} (x,y)= G^{-1}(x-y)\, .
\end{equation}
Further, we will  only consider matrices $G$ of the form
\begin{equation}
G^{ab}_{ij}(x-y)=\delta^{ab}\delta_{ij}G(x-y)\, .
\label{an1}
\end{equation}
This form is certainly the right one in the perturbative regime.
In the leading order in perturbation theory, the non-abelian
character of the gauge group is not important, and the integration
in eq.~(\ref{ansatz}) is basically over the $U(1)^{N^2-1}$ group.
The $\delta^{ab}$ structure is then obvious --- there is a
complete democracy between different components of the vector
potential. The $\delta_{ij}$ structure arises in the following
way. If not for the integration over the group, $G^{-1}_{ij}$
would be precisely the (equal time) propagator of the electric
field. However, due to the integration over the group, the actual
propagator is the transverse part of $G^{-1}$. It is easy to check
that the longitudinal part $\partial_iG_{ij}^{-1}$ drops out of
all physical quantities. At the perturbative level, therefore, one
can take $G_{ij}\sim \delta_{ij}$ without any loss of generality.
Outside the perturbative framework eq.~(\ref{an1}) is a genuine
restriction on the Ansatz, and we will adopt this form of the
matrix $G$ in order to simplify our variational calculation.

We can use additional perturbative information to further restrict
the form of $G$. The theory of interest is asymptotically free.
This means that the short distance asymptotics of correlation
functions must be the same as in perturbation theory. Since
$G^{-1}$ in perturbation theory is directly related to correlation
functions of gauge invariant quantities (e.g. $E^2$), we must have
\begin{equation}
G^{-1}(x)\rightarrow  \frac{1}{x^4}\, , \quad x \rightarrow 0\, .
\label{an2}
\end{equation}

Finally, we expect the theory non-perturbatively to have a gap.
In other words, the correlation functions should decay to zero at some distance scale
\begin{equation}
G(x)\sim 0\, , \quad x>\frac{1}{M}\, .
\label{an3}
\end{equation}
We will build this into our variational Ansatz in  a fairly naive way.
We  will take $M$ to be our only variational parameter.
This can be done by choosing for $G(x)$ a particular form that has the UV and IR asymptotics
given by  eqs.~(\ref{an2}) and  (\ref{an3}), like, for example a massive scalar propagator with mass $M$. We find another parameterization slightly more convenient. The form that will be used throughout
this calculation has the following Fourier transform:
\begin{equation}
G^{-1}(k) = \left\{ \begin{array}{ll} \sqrt{ k^{2} ~} &
\mbox{ if  $ k^2>M^2$}\\
 M &  \mbox{ if $k^2<M^2$}
\end{array}
\right.\, .
\label{an4}
\end{equation}
Using a massive propagator instead, changes the results very
little. Eq.~(\ref{ansatz}), together with
eqs.~(\ref{an1},\ref{an4}), defines our  variational Ansatz.

We now have to calculate the energy expectation  value in these
states and minimize it with respect to the only variational
parameter left --- the scale $M$. Note that the perturbative
vacuum is included in this set of states and corresponds to $M=0$.
A non-zero result for $M$ would therefore mean a non-perturbative
dynamical scale generation in the Yang - Mills vacuum. Now we have
to face up to the question of how to calculate averages in the
state eq.~(\ref{ansatz}).

\subsection{The effective $\sigma$-model}

The expectation value of an arbitrary gauge invariant operator ${\mathcal O}$ is given by the functional integral
\begin{align}
\langle {\mathcal O}\rangle
&= \frac{1}{Z}\int DU DU' \langle {\mathcal O}\rangle_{A}\, ,\nonumber\\
\langle {\mathcal O}\rangle_{A}&=
\int DA\, e^{-\frac{1}{2}
\int_{x,y}  A_i^{Ua}(x)G^{-1}(x-y)A_i^{Ua}(y)} \nonumber\\
&\hspace{4cm}\cdot  {\mathcal O}\,
e^{-\frac{1}{2}\int_{x',y'} A_j^{U'b}(x')G^{-1}(x'-y')A_j^{U'b}(y')}\, ,
\end{align}
where $Z$ is the norm of the trial state.
Two simplifications are immediately obvious.
First, for gauge invariant operators ${\mathcal O}$, one of the group integrations is redundant.
Performing the change of variables $A\rightarrow A^U$ (and remembering that both integration measures $DU$ and $DA$ are group invariant), we obtain (omitting the volume of $SU(N)$ factor $\int dU$)
\begin{align}
\langle {\mathcal O}\rangle
&= \frac{1}{Z}\int DU  \langle {\mathcal O}\rangle_{A}\, ,\nonumber\\
\langle {\mathcal O}\rangle_{A}&=
\int DA\, e^{-\frac{1}{2}
\int_{x,y}  A_i^{Ua}(x)G^{-1}(x-y)A_i^{Ua}(y)} \nonumber\\
&\hspace{4cm}\cdot  {\mathcal O}\,
e^{-\frac{1}{2}\int_{x',y'} A_j^{b}(x')G^{-1}(x'-y')A_j^{b}(y')}\, .
\end{align}
Also, since the gauge transform --- eq.~(\ref{gtqcd}) --- of a vector potential  is a linear function of $A$, for fixed $U(x)$ this is a Gaussian integral, and can therefore be performed explicitly for any reasonable
operator ${\mathcal O}$.
We are then left only with a path integral over one group variable $U(x)$. But this is not easy.

Let us consider first the normalization factor $Z$.
After integrating over the vector potential we obtain
\begin{equation}
Z=\int DU\exp\{-\Gamma[U]\}
\label{sigma}
\end{equation}
with
\begin{equation}
\Gamma[U]=\frac{1}{2} \TR \ln{\mathcal M}
+\frac{1}{2}\lambda[G+SGS^T]^{-1}\lambda\, ,
\label{action}
\end{equation}
where multiplication is understood as the matrix multiplication with indices: colour $a$, space $i$ and position (the values of space  coordinates) $x$, i.e.
\begin{gather}
(A B)_{ik}^{ac}(x,z) = \int_{y} A_{ij}^{ab}(x,y) B_{jk}^{bc}(y,z)\, ,\nonumber\\
\lambda O \lambda = \int_{x,y} \lambda^{a}_{i}(x) O^{ij}_{ab}(x-y) \lambda^{b}_{j}(y)\, .
\end{gather}
The trace $\TR$ is understood as a trace over  all three types of indices.
In eq.~(\ref{action})  we have defined
\begin{align}
S^{ab}_{ij}(x,y)&=S^{ab}(x)\delta_{ij}\delta(x-y)\, ,\nonumber\\
{\mathcal  M}^{ab}_{ij}(x,y)&=
[S^{Tac}(x)S^{cb}(y)+\delta^{ab}]G^{-1}(x-y)\delta_{ij}
\label{def1}
\end{align}
where  $S^{ab}(x)=\frac{1}{2} \tr\left(\tau^aU^\dagger\tau^bU\right)$ and
$~\lambda_i^a(x)=\frac{i}{g} \tr\left(\tau^aU^\dagger \partial_iU\right)$ were defined in eq.~(\ref{defin}) and $\tr$ is a trace over colour indices only.
Using the completeness condition for $SU(N)$
\begin{equation}
\tau^{a}_{ij}\tau^{a}_{kl} = 2 \left(\delta_{il}\delta_{jk} -
 \frac{1}{N}\delta_{ij}\delta_{kl}\right)
\label{completeness}
\end{equation}
one can see that $S^{ab}$ is an orthogonal matrix
\begin{equation}
S^{ab}S^{cb} = \frac{1}{4} \tau^{b}_{ij}\tau^{b}_{kl}
\left(U\tau^aU^\dagger\right)_{ji}
\left(U\tau^cU^\dagger\right)_{kl} = \frac{1}{2}\tr\left(\tau^{a}
\tau^{b}\right) = \delta^{ab}\, . \label{orthogonal}
\end{equation}

We have written eq.~(\ref{action}) in a form which suggests a convenient way of thinking about the problem.
The functional integral eq.~(\ref{sigma}) defines a partition function of a non-linear sigma model with the target space $SU(N)/Z_N$ in three dimensional Euclidean space.
The fact that the target space is $SU(N)/Z_N$ rather than $SU(N)$ follows
from the observation that the action eq.~(\ref{action})  is invariant
under local transformations belonging to the centre of $SU(N)$.
This can be trivially traced back to invariance of $A_i^a$ under
gauge transformations that belong to the centre of the gauge
group.

Note that the quantity $U(x)$ has a well defined gauge
invariant meaning, and it is {\it not} itself a matrix of
a gauge transformation. A contribution of a given $U(x)$ to
the partition function eq.~(\ref{sigma}) and to other expectation
values corresponds to the contribution to the same quantity
from the off-diagonal matrix element between the initial
Gaussian and the Gaussian gauge rotated by $U(x)$.
Consequently,
if matrices $U(x)$ which are far from unity give significant
contribution to the partition function, it means that the off-diagonal contribution is large, and therefore that the simpleminded non gauge invariant Gaussian approximation
(which neglects the off diagonal elements) misses important
physics.

The action of this sigma model is rather complicated.
It is a non-local and a non-polynomial functional of $U(x)$.
There are, however, two observations that will help us devise an
approximation scheme to deal with the problem. First, remembering
that the bare coupling constant of the Yang-Mills theory is small,
let us see how it enters the sigma model action. It is easy to
observe that the only place it enters is in the second term in
 the action eq.~(\ref{action}), because $\lambda^{a}_{i}(x)$ has an explicit
 factor $1/g$.
Moreover, it enters in the same way as a
coupling constant in a standard sigma model action. We can, therefore, easily set up perturbation theory in our sigma model.
With the standard parameterization
\begin{equation}
U(x)=\exp \left\{i\frac{g}{2}\phi^a\tau^a \right\}
\end{equation}
one gets $\lambda^{a}_{i}(x) = - \partial_{i}\phi^a(x) + O(g),~S^{ab}(x) = \delta^{ab} + O(g)$ and
the leading order term in the action becomes
\begin{equation}
\frac{1}{16}\int_{x,y}
\partial_i\phi^a(x)G^{-1}(x-y)\partial_i\phi^a(y) \, .
\end{equation}
This is a free theory albeit with a non-standard propagator which
at large momenta behaves as
\begin{equation}
D(k)\sim G(k) \frac{1}{k^{2}} \sim \frac{1}{|k|^3} \, .
\label{prop}
\end{equation}
It is easy to see that in this sigma model perturbation theory
the coupling constant renormalizes logarithmically. The first
order diagram that contributes to the coupling constant
renormalization is the tadpole. In a sigma model
with a standard kinetic term this diagram diverges linearly
 as $\int d^{3}k/k^{2}$,
a sign of perturbative non-renormalizability. In our model,
however, due to a non-standard form of the kinetic term
eq.~(\ref{prop}), the diagram diverges only logarithmically as
  $\int d^{3}k/k^{3}$.
The form of the $\beta$-function is, therefore, very similar
to the $\beta$-function in ordinary QCD perturbation
theory.
It is a straightforward albeit tedious matter to calculate the one loop graphs in the $\sigma$-model
perturbation theory
 and to extract the renormalization of the coupling constant \cite{Brown:1997gm}.
The result is
\begin{equation}
\beta(g)=-\frac{g^3}{(4\pi)^2}4N \, .
\label{beta}
\end{equation}
For comparison, the pure Yang-Mills $\beta$-function is
\begin{equation}
\beta(g)=-\frac{g^3}{(4\pi)^2}\left(4-\frac{1}{3}\right)N \, .
\end{equation}
The two almost coincide.
The first contribution to the Yang-Mills $\beta$-function
which is reproduced by eq.~(\ref{beta}) is due to the longitudinal gluons, or in other words
to the implementation of  Gauss' law. Since we have implemented Gauss' law exactly on our
trial wave function, this (major) part of the $\beta$-function is correctly reproduced by the $\sigma$-model
renormalization group. The second contribution, which is not present in eq.~(\ref{beta}), is due
to the dynamics of transverse gluons. The fact that this contribution is missing in the $\sigma$-model
suggests that our Ansatz is not perfect in the ultraviolet. However, as this contribution is relatively
small, we will not be discouraged at this stage. We will simply think of eq.~(\ref{beta}) as representing the
complete one loop Yang-Mills $\beta$-function, keeping in mind that it would indeed be  very
interesting to eliminate
this discrepancy by perhaps exploring a less simplistic form of the variational propagator $G(k)$ \cite{Diakonov:1998ir}.

Since the $\sigma$-model is asymptotically free,  perturbation theory becomes worse and worse
as we go to lower momenta, and at some point becomes inapplicable.

Now, however, let us look at the other side of the coin:
let us see how the action looks for the matrices
$U(x)$ which are slowly varying in space.
Due to the short
range of $G(x)$, for $U(x)$ which contain only
momenta lower than the variational scale $M$ the action is
local.
In fact, with our  Ansatz  eq.~(\ref{an4}), it becomes the
standard local action of the non-linear $\sigma$-model
\begin{equation}
\Gamma_L[U] =\frac{M}{2g^2} \tr \int_{x}  \partial_i
U^\dagger(x)\partial_iU(x) + \ldots\label{ac}
\end{equation}
 In this low-momentum approximation we also neglected  the
 space dependence of $S^{ab}_{ij}(x)$ in the term $SGS^T$ in
 eq.~(\ref{action}); then, using the fact that $S$ is an orthogonal
 matrix eq.~(\ref{orthogonal}), one gets $SGS^T \rightarrow G$.

Strictly speaking, due to the $Z_N$ local symmetry of eq.~(\ref{action}), the action for the low
momentum modes is slightly different. The derivatives should
be understood as $Z_N$ covariant derivatives. The most convenient
way to write this action would be to understand $U(x)$ as
belonging to $U(N)$ rather than $SU(N)$ and introduce a $U(1)$
gauge field by
\begin{equation}
\Gamma_L={1\over 2}{M \over g^2} \tr
 \int_x (\partial_i-ib_i)U^\dagger(x)(\partial_i+ib_i)U(x)\, .
\label{low}
\end{equation}
This defines
a sigma model on the target space $U(N)/U(1)$,
which is isomorphic to $SU(N)/Z_N$. This subtlety is unimportant for large $N$ and will not play a
significant role in our analysis.

The action eq.~(\ref{ac}) does not look too bad. Even though it
still cannot be solved exactly, it is amenable to analysis by
standard methods, such as the mean field approximation, which in 3
dimensions and for large number of fields should give reasonably
reliable results.

We adopt therefore the following strategy for dealing with the integration over the $SU(N)$ group.
We integrate
perturbatively the high momentum modes of the field $U(x)$.
This is the renormalization group (RG) transformation. We would
like to integrate out all modes with momenta $k^2>M^2$. This
procedure will necessarily generate a {\it local} effective
action for the low momentum modes. At the same time,
 because of the (presumed) equivalence of the RG flows in QCD
 and our effective sigma-model,  the effective coupling constant
 will be the the  running QCD coupling constant $\alpha_{QCD}(M)$
 at scale $M$. This part of the theory can
then be solved in the mean field approximation.
Clearly, in order for the perturbative RG transformation to be
justified, the QCD running coupling constant at the scale $M$
must be small enough. Our procedure makes sense provided
the energy is minimized at a value of the variational
parameter $M$ for which
\begin{equation}
\alpha_{QCD}(M)<1\, .
\label{consistency}
\end{equation}
We will check whether this consistency condition is satisfied at
the end of the calculation.
In the next section we will calculate the expectation value of
the Hamiltonian in the lowest order of this approximation scheme,
and perform the minimization with respect to $M$.

\subsection{Solving the variational equations -- dynamical mass generation}

We will now calculate the energy, i.e. the expectation value of
the Hamiltonian eq.~(\ref{ymham}). We first perform the Gaussian
integrals over the vector potential at fixed $U(x)$. Let us
consider, for example, the calculation of
 the chromoelectric energy:
\begin{align}
\int_x \langle E_i^{a2} \rangle_{A}
&= \int_x
\left\langle - \frac{\delta}{\delta A_{i}^{a}(x)}
\frac{\delta}{\delta A_{i}^{a}(x)}\right\rangle_{A} \nonumber \\
&=\TR\, G^{-1} - \int_{x,y,z}
G^{-1}(x-y)  G^{-1}(x-z) \langle A_{i}^{a}(y) A_{i}^{a}(z)\rangle_{A}\, .
\end{align}
The Gaussian averaging over $A$ is easily performed. Defining for
convenience
\begin{equation}
a_i^a(x)=
\int_{y,z}
\lambda_i^b(y)G^{-1}(y-z)S^{bc}(z)({\mathcal M}^{-1})^{ca}(z,x)
\label{a}
\end{equation}
one gets
\begin{align}
\int_x \langle E_{i}^{a2}\rangle_{A}  =
3(N^2-1)\int_x  G^{-1}(x,x)
&-\int_x (G^{-1}{\mathcal M}^{-1}G^{-1})^{aa}_{ii}
(x,x)\nonumber\\
&\quad - \int_{x,y}  a_i^a(x)G^{-2}(x-y)a_i^a(y) \, .
\end{align}
 For the chromomagnetic contribution  the calculation is
 straightforward  and one gets
\begin{equation}
\langle (\epsilon_{ijk}\partial_jA^a_k)^2\rangle_{A}
= (\epsilon_{ijk} \partial_ja^a_k)^2
+ \epsilon_{ijk}\epsilon_{ilm} \partial^x_i\partial^y_l
({\mathcal M}^{-1})^{aa}_{km}(x,y)|_{x=y} \, ,
\end{equation}
\begin{multline}
 \langle \partial_jA_k^aA_l^bA_m^c\rangle_{A}
 = \partial_ja_k^aa_l^ba_m^c+
\partial_ja^a_k({\mathcal M}^{-1})^{bc}_{lm}(x,x) \\
 +   a^b_l\partial_j^x({\mathcal M}^{-1})^{ac}_{km}(x,y)|_{x=y}
 +a^c_m\partial_j^x({\mathcal M}^{-1})^{ab}_{kl}(x,y)|_{x=y}\, ,
\end{multline}
and
\begin{align}
\epsilon_{ijk}\epsilon_{ilm}f^{abc}f^{ade}
\langle A_j^bA_k^c A_l^d&A_m^e\rangle_{A} =
2f^{abc}f^{ade}a_j^ba_k^ca_l^da_m^e \nonumber \\
&+ 8f^{abc}f^{ade}a_i^ba_i^d ({\mathcal M}^{-1})^{ce}(x,x)\nonumber\\
&+12f^{abc}f^{ade}({\mathcal M}^{-1})^{bd}(x,x)\, .
({\mathcal M}^{-1})^{ce}(x,x)
\end{align}
Here, we have used the obvious notation
${\mathcal M}^{ab}_{ij}={\mathcal M}^{ab}\delta_{ij}$.
The next step is to decompose the matrix field $U(x)$
into low and high momentum modes. In general this is a
non-trivial problem. However, since we are only going to
integrate over the high momenta in the lowest order in
perturbation theory, for the purposes of our
calculation we can write
\begin{equation}
U(x)=U_L(x)U_H(x)
\end{equation}
where $U_L$ contains only modes with momenta $k^2<M^2$, and
$U_H$ has the form $U_H=1+ig\tau^a\phi_H^a$ and $\phi_H$
contains only momenta $k^2>M^2$.
This decomposition is convenient, since it preserves the group
structure. Also, since the measure $DU$ is group invariant,
we can write it as $DU_LDU_H$. With this decomposition
we have:
\begin{equation}
\lambda_i^a(x)=S_H^{ab}(x)\lambda_{iL}^b(x)+\lambda_{iH}^a(x)\, .
\end{equation}
Further simplifications arise, since we only have to keep the
leading piece in $\phi^a_H$. In this approximation:
\begin{align}
S^{ab}(x)&=S^{ab}_L(x) \, ,\nonumber \\
{\mathcal M}^{ab}(x,y)&=2\delta^{ab}G^{-1}(x-y)\, ,\nonumber \\
\lambda_i^a(x)&=\lambda_{iL}^a(x)+\lambda_{iH}^a(x)\, ,\nonumber \\
a^a_i(x)&={1\over 2}\lambda_{iL}^a(x)+{1\over 2}
\lambda_{iH}^b(x)S_L^{ba}(x)\, .
\label{simple}
\end{align}
The chromoelectric part of the energy can then be written
\begin{multline}
\int_x \langle E_{i}^{a2}\rangle_{A}
={3(N^2 - 1)\over 2}\int_{x} G^{-1}(x,x)
- {1\over 4}
\int_{x,y} \lambda_{iL}^a(x)G^{-2}(x-y)\lambda_{iL}^a(y)\\
- {1\over 4}\int_{x,y}
\lambda_{iH}^a(x)G^{-2}(x-y)\lambda_{iH}^a(y)\, .
\label{esq}
\end{multline}
The cross term vanishes since to this order, as we shall
see, there is a decoupling between the high and the low
momentum modes in the action, and therefore the product
factorizes, and also $\langle\lambda_{iH}^a\rangle=0$.
The Ansatz eq.~(\ref{an4}) allows us to simplify
this expression further. Recalling  that $\lambda_L(x)$ contains
only momenta below $M$, it is immediate to see that
\begin{equation}
\int_{x,y}\lambda_{iL}^aG^{-2}(x-y)
\lambda_{iL}^a(y)=M^2\int_x \lambda_{iL}^a(x)\lambda_{iL}^a(x)\, .
\end{equation}
We can then rewrite eq.~(\ref{esq}) as
\begin{multline}
\int_x \langle E_{i}^{a2}\rangle_{A}  = {3(N^2-1)\over 2}
\int G^{-1}(x,x)  \\
- {M^{2}\over 4}\int_x
\lambda_{iL}^a(x)\lambda_{iL}^a(x)
 - {1\over 4}\int_{x,y}
\lambda_{iH}^a(x)G^{-2}(x-y)\lambda_{iH}^a(y)\, .
\label{esq1}
\end{multline}
The contribution of the magnetic term to the energy is very simple.
All cross terms between the low and high momentum modes drop out.
Some vanish for the same reason as the cross terms in eq.~(\ref{esq}),
and others because they are explicitly multiplied
by a power of the coupling constant. Since our approximation
is the lowest
order in $g$, except for the non-analytic contributions that
come from the low mode effective action, those terms do not
contribute. In fact, the entire low momentum mode contribution
drops out of this term. The reason for this is that the only term which
could give a leading order contribution, i.e.
\begin{equation}
\int (\epsilon_{ijk}\partial_j\lambda^a_{kL})^2
\end{equation}
can be rewritten as
\begin{equation}
(f^{a}_{ijL})^2+O(g^2)\, ,
\end{equation}
where $f^{a}_{ijL}$ is the ``magnetic field" corresponding to
the ``vector potential" $\lambda^a_{iL}$.
However, $\lambda_L$ has the form of a pure gauge vector potential.
Therefore
$f^{a}_{ijL}=0$, and the contribution of this term is higher order
in $g^2$. One can check a posteriori that including this term indeed changes
the energy density in the best variational state by a small amount
($O(10\%)$), but has no effect at all on the best value of the
variational parameter $M$.
The entire magnetic field contribution to the energy is then
\begin{equation}
{1\over 2}\langle B_i^{a2}\rangle_{A}={1\over 8}
 (\epsilon_{ijk}\partial_j
\lambda^a_{kH})^2+{N^2-1\over 2} \partial_i^x\partial_i^yG(x-y)|_{x=y}\, .
\end{equation}
The last step is to perform an averaging over the $U$-field.
For convenience, we rewrite  the complete expression for
the energy density (here $V = \int d^{3}x$ is a space volume)
\begin{align}
{\langle 2H \rangle\over V} =
&{3(N^2-1)\over 2} G^{-1}(x,x) +
(N^2-1) \partial_i^x\partial_i^yG(x-y)|_{x=y} \nonumber \\
&-{1\over 4 V}\int_{x,y} \langle\lambda_{iH}^a(x)G^{-2}(x-y)
\lambda_{iH}^a(y)\rangle_U
+{1\over 4} \langle(\epsilon_{ijk}\partial_j \lambda^a_{kH})^2\rangle_U
\nonumber \\
&-{M^{2}\over 4 V}\int_x
\langle\lambda_{iL}^a(x)\lambda_{iL}^a(x)\rangle_U
\label{energy}
\end{align}
 where  the averaging  over the $U$-field
should be performed with the
sigma model action eq.~(\ref{action}).
In our approximation this action has a simple form.
Using  eq.~(\ref{simple})  we obtain
\begin{equation}
\Gamma ={1\over 4}\int_{x,y} \lambda^a_{iH}(x)G^{-1}(x-y)
\lambda^a_{iH}(y)+
{M\over 4}\int_{x} \lambda^a_{iL}(x)\lambda^a_{iL}(x)\, .
\label{hl}
\end{equation}
The low momentum mode part is precisely equal to $\Gamma_L$ in
eq.~(\ref{low}).
The only difference is that the coupling constant that appears
in this action should be understood as the running coupling
constant at the scale $M$. This obviously is the only $O(1)$
effect of the high momentum modes on the low momentum
effective action
\begin{equation}
\Gamma_L=
{1\over 2}{M\over g^2(M)}
 \tr \int_x (\partial_i-ib_i)U^\dagger(x)(\partial_i+ib_i)U(x)\, .
\label{low1}
\end{equation}

We are now in a position to evaluate the VEV of energy.
The contribution of the high momentum modes is immediately
calculable. Using the parameterization $U_H(x)=1-{i\over 2}g\phi^a\tau^a$,
we find that $\phi^a$ are free fields with the propagator
\begin{equation}
\langle\phi^a(x)\phi^b(y)\rangle=2\delta^{ab}[\partial^x_i\partial^y_i
G^{-1}(x-y)]^{-1}\Big|_{p^2>M^2}\, .
\label{phiphi}
\end{equation}
Also to this order
$\lambda^a_{iH}(x)=\partial_i\phi^a(x)$ and
 therefore $\epsilon_{ijk}\partial_j\lambda^a_{kH}=0$.
Using eq.~(\ref{phiphi}) one finds
\begin{equation}
{1\over 4 }\int_{x,y} \langle\lambda_{iH}^a(x)G^{-2}(x-y)
\lambda_{iH}^a(y)\rangle_U = V {N^2 -1 \over 2} \int_{M}^{\Lambda}
{d^3k\over (2\pi)^3} G^{-1}(k)
\end{equation}
 where  $\Lambda$ is the ultraviolet cutoff,
  and the contribution of the high momentum modes to the energy
(first  two lines in eq.~(\ref{energy})) is
\begin{align}
{2E_0 \over V}&=
(N^2-1)\left\{\int_0^\Lambda {d^3k\over (2\pi)^3}\left[G^{-1}(k)+
k^2G(k)\right]+{1\over 2}\int_0^M{d^3k\over (2\pi)^3}G^{-1}(k)
\right\}\nonumber \\
&=  {N^2-1\over 2\pi^2}\left\{
\int_{0}^{M} k^{2} dk \left[{3\over 2} M + {k^{2} \over M}\right]
 + 2\int_{M}^{\Lambda} k^{3} dk  \right\} \nonumber\\
 &= {N^2-1\over 10\pi^2}M^4+ ...\ .
\label{ehigh}
\end{align}
 Terms omitted in eq.~(\ref{ehigh}) depend on $\Lambda$,
 but are independent of the variational scale $M$.

We now have to evaluate the contribution of the low momentum
modes. It is clear from the form of the action eq.~(\ref{low1})
that this contribution, as a function of $M$, will not be
featureless. The most convenient way to think about it is
from the point of view of classical statistical mechanics.
Comparing eqs.~(\ref{energy}) and (\ref{low1}), we see that
we have to evaluate the internal energy of the sigma model
(with the UV cutoff $M$) at a temperature proportional to
the running coupling constant $g^2(M)$. For large
$M$, the coupling constant is small,
which corresponds to the low temperature regime of the
sigma model. In this regime the global $SU(N)\otimes SU(N)$
symmetry group of the model is spontaneously broken. Lowering
$M$, we raise $g^2(M)$, and therefore the temperature. At some
critical value $g_C$, the $\sigma$-model undergoes a phase transition
into the unbroken (disordered) phase. Clearly, in the vicinity
of the phase transition all thermodynamical quantities will
vary rapidly, and therefore this is a potentially interesting
region of coupling constants.

Before analysing the phase transition region let us calculate
$E(M)$ for large $M$. In this regime the low momentum theory is
weekly coupled. The calculation is straightforward, and to
lowest order in $g^2$ gives
\begin{equation}
{1\over 4}M^2\langle\lambda^a_{iL}(x)\lambda^a_{iL}(x)\rangle=
{N^2-1\over 12\pi^2}M^4\, .
\label{elow}
\end{equation}
Putting this together with the high momentum contribution, we find
\begin{equation}
{E(M)\over V}= {N^2-1\over 120\pi^2}M^4, \qquad
 M\gg\Lambda_{QCD}\, .
\label{largeM}
\end{equation}
This indeed is the expected result. The energy density monotonically
increases as $M^4$, with a slope which is given by the standard
perturbative expression. Note, however, that the slope is very
small, and the contribution of the low momentum modes to the
energy is negative. Therefore, if the internal energy of the
sigma model grows significantly in the phase transition region,
the sign of $E(M)$ could be reversed\,\footnote{The energy, of
course, never becomes negative, since eq.~(\ref{ehigh})
 contains  a divergent $M$-independent piece. Here we concentrate only
on the $M$-dependence of $E$.}  and the energy will then be
minimized for $M$ in this region.

To see, whether this indeed happens, we will now study the low
momentum sigma model in the mean field approximation.
We rewrite the partition function by introducing a (hermitian
matrix) auxiliary field
$\sigma$ which imposes a unitarity constraint on $U(x)$
\begin{equation}
Z=\int DU D\sigma Db_i \exp\left(- \Gamma[U,b,\sigma]\right)\, ,
\end{equation}
with
\begin{multline}
\Gamma[U,b,\sigma] = {M\over 2g^2(M)}
 \tr \int_x \left[\left(\partial_i-ib_i\right)
U^\dagger(x)\left(\partial_i+ib_i\right)U(x)\right. \\
+ \sigma \left. \left(U^\dagger U-1\right)\right] \, .
\end{multline}
The role of the vector field $b_i$ is to impose a $U(1)$ gauge
invariance and, thereby, to eliminate one degree of freedom. As
far as the thermodynamical properties are concerned, its effect is
only felt as an $O(1/N^2)$ correction. At the level of accuracy of
the mean field approximation, we can safely disregard it, which we
do in the following. The mean field equations are:
\begin{equation}
\langle U^\dagger U\rangle=1\, ,
\label{one}
\end{equation}
\begin{equation}
\langle\sigma U\rangle=0\, .
\label{two}
\end{equation}
From eq.~(\ref{two}) it follows that either $\langle\sigma\rangle=0$,
$\langle U\rangle\ne 0$ (the ordered, broken symmetry phase with massless
Goldstone bosons),  or $\langle\sigma\rangle\ne0$, $\langle U\rangle= 0$
(the disordered, unbroken phase with massive excitations).
We are mostly interested in the disordered phase, since there the mean
field approximation should be reliable. Since the symmetry is
unbroken, the expectation value of $\sigma$ should be proportional to
 a unit matrix
\begin{equation}
\langle\sigma_{\alpha\beta}\rangle=\sigma^2 1_{\alpha\beta}\, .
\end{equation}
Eq.~(\ref{one}) then becomes
\begin{multline}
2N^2{g^2(M)\over M}\int_0^M{d^3k\over (2\pi)^3}{1\over k^2+\sigma^2}\\
=
{N^{2} g^2(M)\over \pi^2} \left(1 - {\sigma\over M}
\arctan{M\over \sigma}\right) = N\, .
\label{gap}
\end{multline}
The gap equation, eq.~(\ref{gap}), has solution only for couplings
 (temperatures)  $g^{2}(M)$
 larger than the
 critical coupling (temperature) $g^{2}_{C}$, which
is determined by the
condition that $\sigma = 0$
\begin{equation}
\alpha_C={g^2_C\over 4\pi}={\pi \over 4}{1\over N}\, .
\label{gc}
\end{equation}
The low momentum mode contribution to the ground state energy is
\begin{equation}
N^2M\int_0^M{d^3k\over (2\pi)^3}{k^2\over k^2+\sigma^2}=
{N^2\over 2\pi^2}M\left[{1\over 3}M^3-\sigma^2M+\sigma^3
\arctan{M\over \sigma}\right]\, .
\end{equation}

The final mean field expression for the ground state energy
density is\,\footnote{We do not distinguish between $N^{2}$ and
$N^{2} - 1$ since we have neglected the contribution of the $U(1)$
gauge field. The errors are of order $1/N^{2}$ and even for $N=2$
are very likely smaller than the error introduced by using the
mean field approximation in the first place.}
\begin{equation}
E= {N^2\over 4\pi^2}M^4\left[-{2\over 15}+{\sigma^2\over M^2}
{\alpha_C\over \alpha(M)}\right]
\label{fen}
\end{equation}
where $\alpha(M)$ is the QCD coupling at the scale $M$, $\alpha_C$
is given by eq.~(\ref{gc}), and $\sigma$ is determined by
\begin{equation}
{\sigma\over M}\arctan {M\over \sigma}={\alpha(M)-\alpha_C\over \alpha(M)}\, .
\end{equation}

\begin{figure}
\epsfxsize=7cm
\centerline{\epsfbox{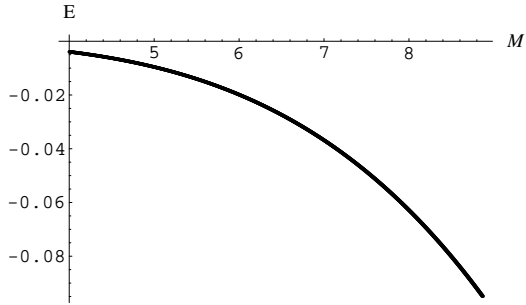}}
\caption{Energy density of a variational trial state as a function of the variational parameter $M$ in units of $\Lambda_{QCD}$. The energy is only shown for $M<8.86 \Lambda_{QCD}$, which corresponds to the disordered phase of the effective low momentum $\sigma$-model.}
\label{fig:energy}
\end{figure}

The energy as a function of $M$  is plotted on Fig.~\ref{fig:energy} for $N=3$.
Qualitatively
it is the same for any $N$.
The minimum of the energy is obviously at the point
$\alpha(M)=\alpha_C$.
Using the one-loop Yang-Mills $\beta$ function and
$\Lambda_{QCD}=150$\,Mev, we find for $N=3$
\begin{equation}
M=\Lambda_{QCD}\, e^{24\over 11}=8.86\,\Lambda_{QCD}=1.33\, {\rm Gev}\, .
\label{mc}
\end{equation}

We thus find that the best variational state is non-perturbative
and is characterized by a dynamically generated mass scale. To get
a rough idea whether the value of this mass is reasonable we have
calculated the value of the gluon condensate
 $(\alpha/\pi)\langle F_{\mu\nu}^{a}F_{\mu\nu}^{a}\rangle =
 (2\alpha / \pi)\left(\langle B^{a\,2}_{i}\rangle - \langle E^{a\,2}_{i}\rangle\right)$.
These calculations are straightforward, and yield\,\footnote{We
have again kept only the $M$-dependent pieces. Each one of the
quantities $\langle E^{a\,2}_{i}\rangle$ and $\langle
B^{a\,2}_{i}\rangle$ is of course positive, due to positive UV
divergent, but $M$ independent, pieces. It is easy to check that
the energy density
 $ E = 1/2(\langle E^{a\,2}_{i}\rangle + \langle B^{a\,2}_{i}\rangle ) = - (1/30\pi^{2})
 N^{2} M^4$ coincides with the first term in eq.~(\ref{fen}), as it must.}
\begin{equation}
\langle E^{a\,2}_{i}\rangle = -{1 \over 24 \pi^2} N^{2} M^4\, ,\qquad
\langle B^{a\,2}_{i}\rangle = -{1 \over 40 \pi^2} N^{2} M^4\, ,
\end{equation}
 and finally
\begin{equation}
{\alpha\over \pi}\langle F_{\mu\nu}^{a}F_{\mu\nu}^{a}\rangle =
{N\over 120\pi^2}M^4=0.008~ {\rm Gev}^4\, .
\end{equation}

The preferred phenomenological value of this condensate is $0.012$
Gev$^4$ \cite {Shifman:1979bx}, although the uncertainty in this
number is large. Considering this, and the unsophisticated nature
of our calculation, the agreement is quite reasonable.

A natural question is of course whether one can assign to $M$
directly the meaning of some physical mass? In the initial
Gaussian wave function, before the projection, it appears as the
``gluon mass''. However the projection changes the wave functional
very considerably, and the direct meaning of $M$ (apart from it
being a dimensional variational parameter) is not so clear.
Nevertheless, naively one expects that, as the gauge invariant
operators $E^2-B^2$ and $EB$ to leading order in $g^2$ are
quadratic in the gluon field, the scale $1/2M$ should appear as
the correlation length in these correlators. Thus we are tempted
to identify $2M$ with the glueball mass. An attempt to calculate
glueball correlation functions was made in \cite{Gripaios:2002bu}.
The result is somewhat unexpected. It was found in
\cite{Gripaios:2002bu} that the scale $2M$ does indeed dominate
the long distance behaviour of the correlation function of the
pseudoscalar glueball. However it appears that the scalar glueball
correlator at long distances is dominated by the scale $\sigma$
--- the gap of the effective $\sigma$-model. If correct, this would mean that the phenomenologically
acceptable scalar glueball mass can only arise in our calculation
if the phase transition in the effective $\sigma$-model is of the
first order with significant latent heat (see next subsection).
This point clearly warrants further investigation.

\subsection{Is it safe?}

For $N=3$, the value of the $QCD$ coupling constant at the
variational scale is $\alpha_C=0.26$. It is reasonably small, so
that the consistency condition for the perturbative integration of
the high momentum modes is satisfied. However, it is not so small
that higher order corrections would be negligible. We expect
therefore that including higher orders in perturbation theory
could give corrections to our result for $\alpha(M)$ of order
$25\%$. Since $M$ depends exponentially on $\alpha(M)$, such
change in $\alpha$ may change the value of $M$ by a factor of
$2-3$. Consequently our result for the dynamical scale $M$ and
other dimensional quantities  should be taken only as an order of
magnitude estimate. In particular, the value of the condensate
$\langle F^2\rangle$ is proportional to the fourth power of $M$,
and would change dramatically as a result of a moderate change of
$\alpha_c$. This is the normal state of affairs in theories with a
logarithmically running coupling constant.
 The best accuracy is always achieved for dimensionless
quantities, since those usually
are slowly varying functions of $\alpha$. The overall scale
depends on $\alpha$ exponentially, and therefore always has the
largest error.

The use of the mean field
approximation to analyze the $\sigma$-model introduces uncertainties into the result.
As a rule, the mean field approximation gives a good estimate of the
critical temperature. Sometimes, however, it gives wrong predictions
for the order of the phase transition. We believe that this is indeed
the case here. The mean field approximation indicates that the phase transition
is second order. The mass gap in the sigma model vanishes continuously
at the critical point. The universality class describing the symmetry
breaking pattern $\left(SU(N)\otimes SU(N)\right)/ SU(N)$ was
considered in
the context of finite temperature chiral phase transition in QCD.
The results of $\epsilon$ expansion \cite{Pisarski:1984ms} and also numerical
simulations \cite{Brown:1990ev,Brown:1990by} strongly suggest that the phase transition
is of  first order except for $N=2$. In our case there is an additional $Z_N$ symmetry in
the game. However its presence is likely only to
increase the latent heat rather than turn the transition into a
second order one. The reason is that the $Z_N$ gauge invariant
theory allows for the  existence of topological defects --- the $Z_N$ strings ---
and condensation of topological defects frequently leads to
discontinuous phase transitions.

Nevertheless, we believe that the bulk of our results is robust against this
uncertainty.
The mean field approximation should be reliable in the regime where
the mass gap in the sigma model is not too small. At the point $M=4.5
\Lambda_{QCD}$ we find
\begin{equation}
\sigma=0.23 M,  \ \ \ \ \alpha(M)=0.38\, .
\end{equation}
Since the gap of the $\sigma$-model is at this point of the order of the UV cutoff, the mean field
approximation should be  reliable here. Perturbation theory is also still reasonable at this value of $\alpha$.
The fact that the energy is negative and has a
minimum for some $\alpha(M)<0.38$, seems to be, therefore, unambiguous.

In fact, independently of the mean field calculation,
it is physically very plausible that the energy is minimized precisely
at the critical temperature, on the disordered side of the phase
transition (if it is of the  first order).
Consider first, the contribution of the high momentum modes to the
ground state energy, eq.~(\ref{ehigh}). It is proportional to $M^4$
with a fixed ($M$-independent) proportionality coefficient
$x= (N^2-1)/ 10\pi^2 $.  Consider now the low momentum contribution
in the large $M$ region, eq.~(\ref{elow}).
 It is again proportional to
$M^4$ with the coefficient $y_0=N^2/12\pi^2$. The proportionality
coefficient of the low momentum contribution at the phase
transition point, according to our calculation, is twice as big
$y_C= 2N^2/12\pi^2 $. This is physically quite transparent. In the
large $M$ -- low temperature regime the global $SU(N)\otimes
SU(N)$ symmetry of the sigma model is broken down spontaneously to
$SU(N)$. This leads to the appearance of $N^2-1$ massless
Goldstone bosons. At zero temperature, those are the only
propagating degrees of freedom in the model. All the rest have
masses of the order of UV cutoff, and therefore do not contribute
to the internal energy. When the temperature is raised ($M$ is
lowered), the Goldstone bosons remain massless and other
excitations become lighter. If the transition is second order, at
the phase transition point the symmetry is restored, and one
should have  a complete multiplet of the $SU(N)\otimes SU(N)$
symmetry of massless particles. The dimensionality of this
multiplet is $2N^2$. The contribution of every degree of freedom
to the internal energy is still roughly the same as at zero
temperature. This is so, since, although at the phase transition
the particles are interacting, critical exponents of scalar
theories in 3 dimensions are generally very close to their values
in a free theory \cite{Brezin:1973jc}. The internal energy of the
$\sigma$-model at this point therefore should be roughly twice its
value at zero temperature. Moving now to higher temperatures, all
the particles become heavier, and therefore their contribution to
internal energy decreases. The internal energy therefore should
have a maximum at the phase transition temperature.

Note that the ground state energy of the Yang-Mills theory  is the
difference between the high momentum contributions and the
internal energy of the low mode sigma model. Already at zero
temperature, these two contributions differ only by $20\%$, which
is why the coefficient in the expression eq.~(\ref{largeM}), even
though positive, is so small. At the critical point, where the low
momentum mode internal energy is twice as large, the chances of
the slope becoming negative are very good. This is what happens in
our mean field analysis but, according to the previous argument,
this in large measure is independent of the approximation. If the
phase transition is first order one should be more careful. The
internal energy then changes discontinuously across the phase
transition. The particles in the disordered phase are always
massive, and the internal energy is smaller than in the case of
the second order phase transition. However, if the transition is
only weakly first order the same argument still holds (the fact
that the mean field predicts second order phase transition may be
an indication that if it is in fact first order, it is only weakly
so). It does seem quite likely that the Yang-Mills ground state
energy will become negative, since it only needs the internal
energy of the $\sigma$-model to grow by $20\%$ at the phase
transition relative to its zero temperature limit. In this case
there will be a finite latent heat, which means that the internal
energy in the disordered (high temperature) phase is higher. The
Yang-Mills variational ground state energy, therefore, will have
its minimum in the disordered phase.

There is good reason to believe, therefore, that these results are qualitatively correct,
and will survive the improvement of the approximation.

\subsection{Instantons}

Instantons are believed to play a very important role in the non-perturbative dynamics of QCD \cite{D95}.
It is, therefore, interesting
to see if our variational calculation has any relation to the instanton approach.

The first thing is to
understand how  we expect to see instantons in this formalism.
Instantons are localized, finite-action classical solutions of the
field equations of QCD in Euclidean space-time.
Physically they represent tunnelling processes between
topologically distinct  vacuum sectors with the exponent of the
instanton action being equal to the transition probability between two
of these vacuum states.
Although the notion of the instanton is intrinsically Euclidean, the
tunnelling between different vacuum sectors can be formulated both in the
Hamiltonian and the Lagrangian languages.  In fact, the
gauge projected  variational approach we discuss here is very well suited for this
purpose. The projection of the initial Gaussian onto the gauge
invariant subspace is achieved by the integration over the gauge group.
The effective
$\sigma$-model arises as an integration over the relative gauge
transformations between the two Gaussian states in the linear
superposition eq.~(\ref{ansatz}). The Boltzmann factor $\exp(-\Gamma[U])$ for
a given matrix $U$ is therefore just the overlap of the initial and the
gauge rotated state or, in other words, the transition amplitude between
the two states\,\footnote{Since the space of the matrices $U$ is
continuous, strictly speaking the Boltzmann factor is the differential
rather than the total amplitude.}.  The instanton transition is
precisely a transition of this type, where the two states (at $t\rightarrow-\infty$ and $t\rightarrow+\infty$)
are related
by a large gauge transformation.  The matrix of this large gauge
transformation must carry a non-zero topological charge $\Pi_3
(SU(N))$.

The integration measure over $U$ indeed includes integration over
topologically non-trivial configurations.  The finiteness of the action
eq.~({\ref{action}) requires that the matrix $U$ approaches a constant
value at infinity.  This identifies all points at spatial infinity,
hence the physical space of the model is $S^3$.  Field configurations
are maps from $S^3$ into the manifold of $SU(N)$ and are classified by
their winding number, or topological charge, which is an element of the
homotopy group $\Pi_3 (SU(N)) = Z$.  The $\sigma$-model action in a
given topological sector is minimized on some configuration which is a
solution of classical $\sigma$-model equations of motion. In
particular, the solution with a unit topological charge is expected to
have a ``hedgehog'' structure much like the topological soliton in the
Skyrme model\cite{B94}.  The integral over $U$ in the steepest descent
approximation is saturated by these classical solutions.

These $\sigma$-model configurations that belong to a non-trivial
topological sector with a unit winding number represent QCD transitions
between the topologically distinct sectors.  The topologically
non-trivial classical soliton solutions of the $\sigma$-model are
therefore the three dimensional images of the QCD instantons.

The QCD instantons are defined in space-time and are therefore four
dimensional point-like objects. The $\sigma$-model solutions are
intrinsically three dimensional.  Nevertheless, there is a natural
simple relation between the two. For a given Yang-Mills instanton
solution $A^{\rm inst}(x_\mu)$ one can find a three dimensional $SU(N)$
matrix $U(x_i)$ by the procedure discussed by Atiyah and Manton
\cite{Atiyah:1989dq},
\begin{equation}
U_{\rm AM}(x_i)=P\exp \left( i\int_Cdx^\mu A^{\rm inst}_\mu \right)\, ,
\end{equation}
where the contour of integration $C$ is a straight line $x_i=const$,
$-\infty<x_0<\infty$.  The matrix $U_{\rm AM}$ gives the relative gauge
transformation between the initial trivial vacuum at
$x_0\rightarrow-\infty$ and the topologically non-trivial vacuum at
$x_0\rightarrow +\infty$ or, in other words, between the initial and
final states of the instanton transition.  Clearly, its meaning is
precisely the same as that of the classical soliton solution of the
effective $\sigma$-model eq.~(\ref{action}).  Also, the QCD instanton
action and the $\sigma$-model soliton action have the same physical
meaning. They both give the transition probability between different
topological sectors in QCD.  We will therefore refer to the
$\sigma$-model solitons as instantons in the following.

Although the QCD  and the $\sigma$-model
instantons have the same physical meaning,
it is not assured that the numerical value for
their respective actions is the same.
They both approximate the value of the transition probability in QCD,
but the approximations involved are quite different. The QCD instanton
action is the result of the standard WKB approximation which is
valid at weak coupling and therefore for small instantons, but breaks
down for instantons of large size. The $\sigma$-model instanton
action on
the other hand is the value of this transition probability in a
particular Gaussian variational approximation. It is natural to expect
that the variational calculation underestimates the value of the
transition probability at very weak coupling.
The transition probability is given by the
overlap of the ``ground state'' wave functions in two topological
sectors. For simplicity let us consider a quantum mechanical system
with two vacua at $x_\pm$. If the area below the barrier separating the
vacua is large, the standard WKB instanton calculation is applicable. The
wave function of each of the vacua below the barrier has essentially an
exponential fall off $\exp\{i\int^x \sqrt{E-V(x-x_\pm)}\}$. The
instanton calculation is the calculation of the overlap of these
functions.  Our variational calculation corresponds to approximating
the respective ``ground states'' at $x_\pm$ by Gaussian wave functions.
The tails of the Gaussians fall off much faster away from the minimum
than the actual wave function and the overlap is therefore expected
to be smaller.
When the coupling constant is not too small
(or when the area below the barrier is not too large) the overlap between the
two states is no longer determined by the behaviour of the ``tails'' of
the wave functions. In this situation one can expect the Gaussian
approximation to do much better, since the overlap region contributes
significantly to the energy and therefore plays an important role in the
minimization procedure.

In fact, implicitly, the instantons
played a very important role already in the energy minimization described in this section.
As we have seen above, the energy is minimized for the
value of the mass parameter $M$ at which the $\sigma$-model is in the
disordered phase. The transition between ordered and disordered phases
in a statistical mechanical system can usually be described as a
condensation of topological defects. This is a standard description of
the phase transition in the Ising and XY models \cite{Kogut:1979wt}.  In the
$\sigma$-model eq.~(\ref{action}) the relevant topological defects are
none other than the instantons. In this sense the appearance of the
dynamical mass in the best variational state is itself driven by the
condensation of instantons.

Perhaps the most significant difference between the kinks in the
Ising model and the QCD instantons, is that the former have a
fixed size, while the latter come in a variety of sizes. This is a
direct consequence of the dilatation symmetry of the classical
Yang-Mills action. It is not necessary for instantons of all sizes
to condense in order to drive the transition. The naive
expectation therefore is that the large size instantons (larger
than $1/M$) condense, while the smaller ones should still exist as
semi-classical solutions in the effective $\sigma$-model action.

The simple qualitative argument to this effect is the
following. Consider the effective $\sigma$-model action for very large
size instantons. In such a configuration only field modes with small
momentum $k<M$ are present. For these momenta the action
is the standard local
$\sigma$-model where $M$ plays the role of an ultraviolet cutoff
\begin{equation}
\Gamma={1\over 2}{M \over g^2(M)} \tr
 \int_{x}  \partial_iU^\dagger(x)\partial_iU(x)\, .
\label{lowmodes}
\end{equation}
If the large size instantons are stable at all, they should also
be present
as stable solutions in this local action eq.~(\ref{lowmodes}).
However this is not the case as can be easily seen by the standard
Derrick type scaling argument.
Take an arbitrary  configuration $u(x)$ in the instanton sector
and scale all the coordinates
by a common factor $\lambda$. Then obviously
\begin{equation}
\Gamma[u(\lambda x)]=\lambda^{-1} \Gamma[u(x)]\, .
\end{equation}
The dependence of the action on $\lambda$ is monotonic and
is minimized at $\lambda\rightarrow\infty$.
This means that the instantons in the local $\sigma$-model shrink
to the ultraviolet cutoff $1/M$. For instantons smaller than the
inverse
cutoff we cannot use the
local action anymore. However, the behaviour of these small size
instantons is already familiar.
We know that classically they exist at arbitrary size, but that when the running of the
coupling is taken into account, these instantons are pushed to
the large size. This is the familiar infrared problem of large instantons.
In our variational state, the coupling constant stops running at the
scale $M$. The picture is therefore very simple. The small size
instantons are pushed to larger size by the effect of the coupling
constant, while the large size instantons are pushed to smaller size
by the effect of the local $\sigma$-model scaling. We therefore expect that the instanton
size will be stabilized somewhere in the vicinity of
$\rho\sim1/M$.

The behaviour of the instantons in the variational ansatz
eq.~(\ref{ansatz}) was studied in detail in \cite{Brown:1998cp}.
The results are indeed very much in line with the expectations
just outlined. The action of a small size instanton in the
$\sigma$-model was found to be independent of its size (neglecting
the running coupling effects), and numerically equal to
\begin{equation}
\Gamma = 1.96\, \frac{8 \pi^2}{g^2}\, .
\label{prob}
\end{equation}
This is about twice
the value of the instanton action in QCD: $\Gamma_{inst}=\frac{8 \pi^2}{g^2}$.
Thus, as expected, the tunnelling transition amplitude is underestimated in the Gaussian approximation for small instantons.
Interestingly enough, however, the actual configuration of the $\sigma$-model field that minimizes the action in the one
instanton sector was found to be practically indistinguishable from the Atiyah-Manton expression calculated on the QCD
instanton.
This means that even though the
value of the transition probability is underestimated in the Gaussian
approximation, the actual field configurations into which the
tunnelling is most probable are identified correctly --- they are
precisely the same as in the WKB calculation.

As for the large size instantons, when the running of the coupling constant is taken into account their size is stabilized at
about $\rho=(1-1.5)/M$. The uncertainty is to do with the way the running of the coupling constant is modified at $k<M$.
It is in fact interesting to note that this instanton size
is consistent with the average size of the instantons in
the instanton liquid model of \cite{Shuryak:1982ff,Shuryak:1982dp,Diakonov:1984hh,Shuryak:1989fy,Shuryak:1990cx}.
For the case of $SU(2)$, the average instanton size, in units of the gluon
condensate obtained in the instanton liquid model, turns out to be
\cite{Diakonov:1984hh,Shuryak:1989fy,Shuryak:1990cx},
\begin{equation}
\rho \,
\left(
\langle F_{\mu \nu}^a F_{\mu \nu}^a \rangle
\alpha/\pi \right)^{1/4}
\sim 0.4\, .
\end{equation}
In our case, taking the value of the gluon condensate obtained in the
variational approach, we find
\begin{equation}
\rho \, \left(
\langle F_{\mu \nu}^a F_{\mu \nu}^a \rangle \alpha/\pi
\right)^{1/4}
\sim 0.2 - 0.3\, .
\end{equation}
The relation to the instanton liquid model is an interesting question which deserves further study.

We also note that the
 variational Ansatz which has been considered so far corresponds to a
 zero value of the QCD $\theta$-parameter, since we have integrated
over the entire gauge group without any extra phases. As
is well known, the general $\theta$-vacuum is defined as
\begin{equation}
|\theta> = \sum_{n} e^{i n \theta} |n>
\label{theta}
\end{equation}
where $n$ labels the topological sectors in the configuration
 space (space of all potentials $A_{i}^{a}(x)$).  Generalization
 of our trial wave functions
to non-zero $\theta$ is trivial --- all we need to do
 is to insert in eq.~(\ref{ansatz})
 an extra  phase factor in the integrand
\begin{equation}
\exp\left\{i{\theta\over 24\pi^2}\int dx \epsilon_{ijk}
\tr\left[(U^\dagger\partial_iU)
(U^\dagger\partial_jU)(U^\dagger\partial_kU)\right]\right\}\, .
\label{thetafactor}
\end{equation}
The integrand here is a properly normalized topological charge,
and
it takes integer values for topologically non-trivial configurations
$U(x)$, i.e. this factor reproduces the $\exp(in\theta)$ term in
 eq.~(\ref{theta}). This phase factor can be also obtained  if one
 remembers that usually the $\theta$-dependence of the wave functional
 is given by the $\exp\left[i\theta S_{CS}(A) \right]$, where
 $ S_{CS}(A)$ is a Chern-Simons term, which under the gauge
 transformation $U$ transforms as
\begin{equation}
S_{CS}(A^{U}) = S_{CS}(A) +
{1\over 24\pi^2}\int dx \epsilon_{ijk}
\tr\left[(U^\dagger\partial_iU)
(U^\dagger\partial_jU)(U^\dagger\partial_kU)\right]\, ,
\end{equation}
so that integrating over $U$ leads precisely to the phase factor
eq.~(\ref{thetafactor}). The state thus constructed is an
eigenstate of an operator of the large gauge transformation with
eigenvalue $e^{i\theta}$. This modification results in the
addition of the same topological term to the effective action
eq.~(\ref{action}). It is amusing to note that for $\theta=\pi$,
the ``Skyrmions" in the effective theory  will be ``fermions".

While the extension of the variational  calculation to
non-vanishing $\theta$-term is quite straightforward, it has not
been performed thus far.

\subsection{Confinement?}

The most interesting question is of course whether our variational state is confining. In the toy model in 2+1 dimensions
discussed in the previous section, we were able to answer this question by calculating both the expectation value
of the spacial Wilson loop, and the potential between static charges. Unfortunately, in the Yang-Mills theory the calculation
is much more complicated and the answer is not known.
Although there are some arguments that the state is indeed confining (see next section), it has not been proved or disproved by a
direct calculation. The calculation of a potential between static charges {\it \'a la} \cite{Kovner:1998eg} has not been attempted. As for the
calculation of the Wilson loop, some progress has been made in reducing this calculation to the $\sigma$-model level, but no final
result has been obtained.

The difficulty in the calculation of
the Wilson loop
\begin{equation}
W(C) = \left\langle \tr~ P \exp\left(i{g\over 2}\oint_C dx_i A_{i}^{a} \tau^{a}
\right)\right\rangle
\label{wloop}
\end{equation}
is to take into account the $P$-ordering
 of the exponent.
One way of doing so is to introduce
   new degrees of freedom living on the contour $C$ which, after
 quantization, become the $SU(N)$ matrices $\tau^{a}$
 \cite{Diakonov:1989fc}. We briefly describe the construction in the
 case of the $SU(2)$ group ---  the generalization of this construction to
an arbitrary Lie group has been discussed in \cite{Diakonov:1989fc}.

 The construction is based on the observation, made in
 \cite{Polyakov:1988md,Alekseev:1988tj}, that instead of considering the ordered product
 of $\tau^{a}$ matrices one can consider the correlation function
\begin{multline}
\left\langle {\tau^{a}(t_{1})\over 2} {\tau^{b}(t_{2})\over 2}
\ldots {\tau^{c}(t_{k})\over 2}\right\rangle \longrightarrow
\langle n^{a}(t_{1}) n^{b}(t_{2})\ldots n^{c}(t_{k})\rangle \\
= \int Dn(t)\,  n^{a}(t_{1}) n^{b}(t_{2})\ldots n^{c}(t_{k})\\
\cdot \exp\left[i(S+1/2) \int_{\Sigma} d^{2}\xi
\epsilon_{\mu\nu}\epsilon^{abc} n^{a}
\partial_{\mu}n^{b}\partial_{\nu}n^{c}\right]
\end{multline}
where $S$ is the spin of the representation, i.e. for the fundamental
 representation $S = 1/2$; $n^{a}(t)$ is a unit vector
 $n^{a}n^{a} = 1$ living on a contour $C$ ($t$ is a coordinate
 on the contour); and $\Sigma$ is an arbitrary two-dimensional
 surface with the boundary $C = \delta\Sigma$. The
 two-dimensional action
\begin{equation}
S[n] = \int_{\Sigma} d^{2}\xi\,
\epsilon_{\mu\nu}\epsilon^{abc} n^{a}
\partial_{\mu}n^{b}\partial_{\nu}n^{c}
\label{S[n]}
\end{equation}
 depends only on values $n^{a}(t)$
 at the boundary.

It can be shown that the Wilson loop can be rewritten as
\begin{multline}
W(C) = \left\langle\int Dn(t)
\exp\left[i\int_{\Sigma} d^{2}\xi\,
\epsilon_{\mu\nu}\epsilon^{abc} n^{a}
\partial_{\mu}n^{b}\partial_{\nu}n^{c}\right]\right.\\
\cdot \left.\exp\left(
ig\oint_C dx_i\, A_{i}^{a}(x(t)) n^{a}(t)
\right)\right\rangle
\end{multline}
The average over $A_{i}$ can be performed using
eqs.~(\ref{action},\ref{a})
\begin{multline}
\left\langle\exp\left(
ig\oint_C dx_i \, A_{i}^{a}(x(t)) n^{a}(t)
\right)\right\rangle_{A}
= \exp\left(-
ig\oint_C dx_i \, a_{i}^{a}(x(t)) n^{a}(t)
\right) \\
\cdot\exp\left(-{1\over2}\oint_C\oint_Cdt_{1}dt_{2}\, \dot{x}_{i}(t_1)
\dot{y}_{i}(t_2)n^{a}(t_1)n^{b}(t_2)({\mathcal M}^{-1})^{ab}(x,y)
\right)
\end{multline}
where $a_{i}^{a}$  was defined in eq.~(\ref{a}).
The Wilson loop can be calculated as the average over
 two scalar fields: $U(x)$ living in the whole space and
 $n^{a}(\xi)$ living on a two-dimensional surface $\Sigma$
 such that $C = \delta \Sigma$
\begin{multline}
W(C) = \int DU \int Dn \, \exp\left(-\Gamma[U] + iS[n]\right)
\exp\left(-
ig\oint_C dx_i \, a_{i}^{a}(x(t)) n^{a}(t)
\right)  \\
\cdot\exp\left(-{1\over2}\oint_C\oint_Cdt_{1}dt_{2}\, \dot{x}_{i}(t_1)
\dot{y}_{i}(t_2)n^{a}(t_1)n^{b}(t_2)({\mathcal M}^{-1})^{ab}(x,y)
\right)\, .
\label{wloopnu}
\end{multline}
In the infrared limit one can use eq.~(\ref{simple}) to
 simplify eq.~(\ref{wloopnu}) and get
\begin{align}
W(C&) = \int DU_{L} \int Dn
\exp\left(-\Gamma_{L}[U] + iS[n]\right)\nonumber\\
&\cdot\exp\left(-
i{g\over 2}\oint_C dx_i \, \lambda_{i~L}^{a}(x(t)) n^{a}(t)
\right)  \nonumber\\
&\cdot\exp\left(-{1\over4}\oint_C\oint_Cdt_{1}dt_{2}\, \dot{x}_{i}(t_1)
\dot{y}_{i}(t_2)n^{a}(t_1)n^{a}(t_2)G(x-y)
\right) \nonumber\\
&\cdot \int DU_{H}\exp\left(-\Gamma_{H}[U]\right)
\exp\left(-
i{g\over 2}\oint_C dx_i \, \lambda_{i~H}^{b}(x(t))S^{ba}_{L} n^{a}(t)
\right)\, .
\label{wloopnu1}
\end{align}
Integrating over $U_H$ one obtains
\begin{align}
W(C) =
\int &Dn \exp\left(iS[n]\right)\nonumber\\
&\cdot \exp\left(-{1\over2}\oint_C\oint_Cdt_{1}dt_{2}\, \dot{x}_{i}(t_1)
\dot{y}_{i}(t_2)n^{a}(t_1)n^{a}(t_2)G(x-y)
\right)
 \nonumber \\
&\cdot \int DU \exp\left(-\Gamma[U]\right)
\exp\left[
{1\over 2}\oint_C dx_i \, \tr\left(\tau^{a}U^\dagger\partial_iU\right)
 n^{a}(t)\right]
\label{wloopfinal}
\end{align}
where the integration $DU$ is over the low momentum modes only and
$\Gamma[U]$ is the corresponding low momentum action. Since
$G(x-y)$ is short range, the term
\begin{equation}
\exp\left(-{1\over2}\oint_C\oint_Cdt_{1}dt_{2}\, \dot{x}_{i}(t_1)
\dot{y}_{i}(t_2)n^{a}(t_1)n^{a}(t_2)G(x-y)
\right)
\end{equation}
gives only perimeter dependence and  can be  neglected
when calculating the string tension. Now reverting back from $n^a$ to $\tau^a$
we find
\begin{equation}
W(C)=\left\langle \tr P\exp\left({1\over 2}
\oint_C dl_i\, U^\dagger\partial_iU\right)\right\rangle_{U}
\label{wl1}
\end{equation}
where the averaging is performed with the low momentum $\sigma$-model action.
This is reminiscent of the average of the monodromy operator
\begin{equation}
M=\tr P\exp\left(\oint_C dl_i\, U^\dagger\partial_iU\right)
\end{equation}
and one might expect that the result is similar. Since the target
space of the sigma model is ${\mathcal M}=SU(N)/Z_N$, and
$\Pi_1({\mathcal M})=Z_N$, the monodromy can take on values
$\exp(i2\pi n/N)$. It has a natural interpretation in terms of the
topological defects in the sigma model. As mentioned above, the
topology of the $\sigma$-model allows for the existence of $Z_N$
strings (the soliton-instantons discussed in the previous
subsection do not play any special role in the monodromy
calculation). The string creation operator and the operator $M$
satisfy the commutation relations of the t'Hooft algebra
\cite{'tHooft:1978hy}. Therefore, in the presence of a string, the
operator $M$ has expectation value $\exp(i2\pi n/N)$, where $n$ is
the linking number between the loop $C$ and the string. As we have
argued, the sigma model is in the disordered phase and the
disordering can be thought of as the condensation of the
topological defects.  We have thus far discussed skyrmions
(instantons) as the relevant defects, but it is quite plausible
that the $Z_N$ strings are condensed as well. That would mean that
the vacuum of the sigma model has a large number of strings and
also that the fluctuations in this number are large. In this
situation the VEV of $\mathcal M$ must average to zero very
quickly, and for large loops will have an area law. We then may
expect that the Wilson loop will also have an area law $W(C) \sim
\exp\left(-\alpha' A\right)$.

While this argument is not implausible, currently we have no quantitative method of
estimating eq.~(\ref{wl1}).

Even though the question about confining properties of the state remains unanswered,
the results of the variational calculation so far are quite interesting.
It yields the dynamical generation of the scale which is of the right magnitude, a reasonable
value of the gluon condensate
and a neat relation to the instanton physics.
All these results are intrinsically non-perturbative.

Apart from the vacuum structure, there is another mysterious
domain of the QCD physics which is not accessible with
perturbative tools: the deconfining phase transition. We may hope
that the variational method can give us a handle to understanding
the deconfinement physics. In the next section we describe its
application to the Yang Mills theory at finite temperature and the
study of the deconfinement phase transition.

\section{The Yang Mills theory at finite temperature}
\label{sec:fin}

Attempts to understand the nature of the deconfining phase
transition in QCD date back almost 30 years. Since the pioneering
work of Polyakov \cite{Polyakov:1978vu} and Susskind
\cite{Susskind:1979up}, much effort has been made to study the
basic physics as well as the quantitative characteristics of the
transition. The high temperature phase of QCD is widely believed
to resemble an almost free plasma of quarks and gluons. At
asymptotically high temperatures this is confirmed by explicit
perturbative calculations of the free energy \cite{Arnold:1995eb}.
Perturbation theory in its simplest form, however, is valid only
at unrealistically high temperatures. In recent years a different
and promising avenue has been explored. This incorporates
analytical resummation of the effects of the gluon screening mass
into the 3D effective Lagrangian, which is then solved numerically
by 3D lattice gauge theory methods
\cite{Braaten:1995cm,Braaten:1996jr,Kajantie:1997pd,Kajantie:2000iz}.
The results of this approach seem to be in agreement
\cite{Hart:1999dj} with direct 4D lattice gauge theory
calculations \cite{Boyd:1996bx,Datta:1999yu} all the way down to
$2\mathsf{T_{c}}$. Although we are quite advanced in the
understanding of the high temperature phase, the transition region
itself is very poorly understood. This region of temperatures,
$\mathsf{T_{c}}<\mathsf{T}<2\mathsf{T_{c}}$, is of course the most
interesting one, since it is in this region that the transition
between ``hadronic" and ``partonic" degrees of freedom occurs.
Interestingly enough, the numerical results indicate that although
asymptotically the free energy does approach that of the free
partonic plasma, the deviations from the Stefan-Boltzmann law even
at temperatures of order $10 \mathsf{T_{c}}$ are quite sizable, of
order of $15\%$. This is an indication that the interesting
physics of the transition region remains important even at these
high temperatures. The study of the transition region itself is a
complicated and inherently non-perturbative problem.

The purpose of this section is to study the deconfining phase
transition in a pure $SU(N)$ Yang-Mills theory using the
variational approach described in the previous sections suitably
extended to finite temperature \cite{Kogan:2002yr}. We will minimize the relevant
thermodynamic potential at finite temperature, i.e. the Helmholtz free
energy, on a set of gauge invariant density matrices.

\subsection{The variational Ansatz for the density matrix}

The equilibrium state of a quantum mechanical system at finite
temperature is not a pure state, but is described by a mixed
density matrix. Thus in order to extend the variational analysis
to finite temperature we have to generalize our ansatz
eq.~(\ref{ansatz}) so that it includes mixed states. In scalar
theories the Gaussian approximation has a long history of
applications at finite temperature
\cite{Eboli:1988fm,Bardeen:1983st}. We generalize our Ansatz along
the same lines.

We start by considering the density matrices which in the field
basis have Gaussian matrix elements \cite{Kogan:2002yr}
\begin{multline}
    \tilde\varrho[A,A']  = \exp \bigg\{ -\frac{1}{2} \int_{x,y}
            A^{a}_{i}(x)G^{-1ab}_{ij} (x,y)A^{b}_{j}(y)\\
            +A^{'a}_{i}(x)G^{-1ab}_{ij} (x,y)A^{'b}_{j}(y)
-2A^{a}_{i}(x)H^{ab}_{ij} (x,y)A^{'b}_{j}(y)\bigg\}\, .
\label{eq:iniansatz}
\end{multline}

As before, we take the variational functions  diagonal in both
colour and Lorentz indices, and translationally invariant
\begin{eqnarray}
    G^{-1ab}_{ij} (x,y) &= &\delta^{ab} \delta_{ij}
    G^{-1}(x-y)\, ,\\
H^{ab}_{ij} (x,y) &= &\delta^{ab} \delta_{ij}
    H(x-y)\, .\nonumber
\end{eqnarray}
Then
\begin{equation}
    \tilde\varrho[A,A']
    = \exp \bigg\{ -\frac{1}{2} \int_{x, y}
    \bigl(A G^{-1} A +  {A'} G^{-1} {A'}
    -2A H {A'} \bigl) \bigg\}\, .
    \label{tilderho}
\end{equation}
For $H=0$ this density matrix represents a pure state, since it
can be written in the form
\begin{equation}
    \tilde\varrho=|\Psi[A]><\Psi[A]|
\end{equation}
with $\Psi[A]$ a Gaussian wave function, eq.~(\ref{psi0}). At non-zero $H$ the
density matrix is, however, mixed. The magnitude of $H$,
therefore, determines the entropy of this trial density matrix.

We now make an additional simplification in our ansatz. First, we
restrict the functions $G^{-1}(x)$ to the same functional form as
at zero temperature eq.~(\ref{an4}), i.e.
\begin{equation}
G^{-1}(k) = \left\{ \begin{array}{ll} \sqrt{ k^{2} ~} &
\mbox{ if  $ k^2>M^2$}\\
 M &  \mbox{ if $k^2<M^2$}
\end{array}
\right.\, .
\label{G}
\end{equation}
Further, we will take $H(k)$ to
be small and non-vanishing only at low momenta
\begin{eqnarray}
\label{H}
 H(k) = \left\{ \begin{array}{ll} 0  &
\mbox{ if  $ k^2>M^2$}\\ H \ll M&  \mbox{ if $k^2<M^2$}
\end{array}
\right.\, .
\end{eqnarray}

The logic behind this choice of Ansatz is the following. At finite
temperature we expect $H(k)$ to be roughly proportional to the
Bolzmann factor $\exp\{-{\mathsf E}(k)\beta\}$. In our ansatz, the
role of one particle energy is played by the variational function
$G^{-1}(k)$. We will be  interested only in temperatures close to
the phase transition, and those we anticipate to be small,
$\mathsf{T_{c}}\le M$. For those temperatures, one particle modes
with momenta $k\ge M$ are not populated, and we thus can put
$H(k)=0$. For $k\le M$ the Bolzmann factor is non-vanishing, but
small. Further, it depends only very weakly on the value of the
momentum. We will have, of course, to verify a posteriori that our
assumptions about the smallness of $\mathsf{T_{c}}$ and $H$ are
justified.

As before, we  explicitly impose gauge invariance by projecting
$\varrho$ onto the gauge invariant sector
\begin{multline}
    \varrho[A,A'] = \int  D U' D U''  \exp \bigg\{
    -\frac{1}{2} \int_{x,y}
    A^{U'} G^{-1} A^{U'} \\
    +  {A'}^{U''} G^{-1}{A'}^{U''}
    - 2A^{U'} H {A'}^{U''}
    \bigg\}\, ,
    \label{eq:ginvansatz}
\end{multline}
where  $A^U$ is given by eqs.~(\ref{gtqcd},\ref{defin}). One of
the group integrations in eq.~(\ref{eq:ginvansatz}) is redundant,
since we will only calculate the quantities of the form
$\TR\varrho O$, with $O$ being gauge invariant. Our Ansatz for the
density matrix then is
\begin{equation}
    \label{eq:ansatz}
    \varrho[A,A'] = \int  D U \exp \bigg\{
    -\frac{1}{2} \int_x
    A G^{-1} A +  {A'}^U G^{-1}{A'}^U - 2A H {A'}^U
    \bigg\}\, .
\end{equation}
This expression is not explicitly normalized to unity.
Nevertheless, we find it convenient to refer to it as density
matrix while explicitly inserting a normalization factor whenever
necessary. Thus the average of a gauge invariant operator ${\mathcal
O}$ is given by
\begin{align}
\label{eq:expvalue}
    \langle {\mathcal O}\rangle_{A,U}
    & = Z^{-1}\TR (\varrho {\mathcal O}) \nonumber  \\
    &= Z^{-1} \int    D U D A
    \:{\mathcal O}(A,A')\nonumber\\
    &\quad\cdot\exp \bigg\{ -\frac{1}{2} \int_x
    A G^{-1} A
    +  {A'}^U G^{-1}{A'}^U
    - 2A H {A'}^U \bigg\}\Bigg|_{A'=A}\, ,
\end{align}
where $Z$ is the normalization of the trial density matrix
$\varrho$, i.e.
\begin{align}
    Z=\TR\varrho &= \int  D U D A \exp \bigg\{
    -\frac{1}{2} \int_x
    A G^{-1} A +  {A}^U G^{-1}{A}^U
    -2 A H {A}^U \bigg\} \nonumber\\
    &=\int  D U D {\tilde A} \exp \biggl\{
    -\frac{1}{2} \int_x
    {\tilde A}\Delta{\tilde A}
    +\lambda \bigl(
    G^{-1} - \omega \Delta^{-1} \omega^{T} \bigr)
    \lambda\biggr\}
\end{align}
with
\begin{eqnarray}
    {\tilde A} &=& A + \lambda \omega\Delta^{-1}\, , \\
    \Delta &=& 2 G^{-1} \Bigl(1-\frac{HG}{2} (S+S^T)\Bigr)\, , \\
    \omega &=& (G^{-1}S-H)\, .
\end{eqnarray}

The ${\tilde A}$ integration can be performed to yield
\begin{equation}
    \TR\varrho = \int  D U
    \exp \bigg\{
    -\frac{1}{2}
    \lambda \bigl(
    G^{-1} - \omega \Delta^{-1} \omega^{T}
    \bigr) \lambda
    -\frac{3}{2}\TR\ln\frac{\Delta}{2}\bigg\}\, .
\end{equation}

We now adopt the same strategy for treating the high momentum
modes of $U$ as at $\mathsf{T}=0$. Namely, they are integrated
perturbatively to one loop accuracy. The result is the effective
$\sigma$-model for the matrices $U$ with momenta below $M$. The
coupling constant $g$ of this $\sigma$-model gets renormalized
as before according to the one loop Yang-Mills $\beta$-function,
and thus has to be understood as $g(M)$. Additionally, due to
independence of $H$ on momentum, for low momentum modes of $U$ the
function $H(x-y)$ is equivalent to $H\delta^3(x-y)$.

The final approximation has to do with the fact that $H$ is
assumed to be small. For arbitrarily large $H$ the variational
calculation is forbiddingly complicated even with all the above
mentioned simplifications. This is because the gauge projection
renders the calculation of entropy in the general case unfeasible.
However, at small $H$ we only need to calculate the leading term
in entropy. This calculation can indeed be done, and is described
in the following.  Since we are only calculating the leading
order contribution in $H$, we only have to consider corrections to
the $\sigma$-model action of first order in $H$. With this in
mind, the normalization factor becomes
\begin{equation}
    \label{eq:trvarrho}
    \TR\varrho = \int  D U \exp \bigg\{
    -\frac{1}{2}\lambda\Big(
    \frac{G^{-1}}{2} + \frac{H}{4} (S+S^T)
    \Big)\lambda+\frac{3}{4} HG\:\tr (S+S^T)\bigg\}\, .
\end{equation}

\subsection{The effective $\sigma$-model}

Just like at zero temperature, the normalization $Z$ can be
interpreted as the generating functional for a theory defined by
the action ${\mathcal S}(U)$
\begin{equation}
    Z=\TR\varrho =  \int  D U e^{- {\mathcal S}(U)}\, ,
\end{equation}
where
\begin{equation}
    {\mathcal S}(U) = \frac{M}{4} \lambda \lambda
    + \frac{1}{8} \lambda \:H (S+S^T)\:\lambda
    - \frac{1}{4\pi^2} HM^2 \tr S\, .
\end{equation}

We simplify this expression using
\begin{eqnarray}
    \label{eq:ll}
   \lambda \lambda  &=& \frac{2}{g^2}\tr( \partial U\partial U^\dagger)\, , \\
    \label{eq:lr}
  \lambda S^T \lambda &= &\lambda S \lambda
    =- \frac{1}{2 g^2}\tr\Bigl[(U^\dagger \partial U - \partial U^\dagger U)
    (\partial U U^\dagger - U\partial U^\dagger) \Bigr] \, ,\\
    \label{eq:trs}
   \tr S &=& \tr S^T = \tr U^\dagger \tr U -1\, .
\end{eqnarray}

Inserting these into the action we get
\begin{multline}
    \label{eq:action}
    {\mathcal S}(U) ={M\over 2 g^2} \tr( \partial U\partial U^\dagger)
    - \frac{H}{8 g^2}\tr\Big[(U^\dagger \partial U - \partial U^\dagger U)
    (\partial U U^\dagger - U\partial U^\dagger) \Big]\\
    - \frac{1}{4\pi^2} H M^2 \tr U^\dagger \tr U\, ,
\end{multline}
where $U$-independent terms have been dropped.

At this point it is useful to relate our effective $\sigma$-model
with a standard tool used in finite temperature calculations,
namely the effective action for the Polyakov loop. The matrix $U$
plays a similar role to the  Polyakov loop $P$ at finite
temperature
--- the functional integration over $U$ projects out the physical
subspace of the large Hilbert space on which the Hamiltonian of
gluodynamics is defined. The effective $\sigma$-model
eq.~(\ref{eq:action}) therefore is a close analogue of the effective
theory for the low momentum modes of the Polyakov loop variable.
Its status and applicability region are however different from the
usual perturbative effective actions, see e.g.
\cite{KorthalsAltes:1994ca}. The standard effective action is
calculated in perturbation theory and is valid at high
temperature. Our effective action eq.~(\ref{eq:action}) depends on
the variational parameters $M$ and $H$, and in a sense is a
variational effective action. Also due to our restrictions to
small values of $H$, a priori we do not expect it to be valid at
high temperatures but, rather, it should represent correctly the
physics in the phase transition region.

Another important difference is that our effective $\sigma$-model
does not have the local gauge invariance $U(x)\rightarrow
V^\dagger(x)U(x)V(x)$ which is usually associated with the
effective action for the Polyakov loop. The reason for this is
that our setup is different from that of the standard finite
temperature calculation. The way this gauge invariance usually
appears
 is the following. Consider the calculation of any gauge
invariant observable in the equilibrium density matrix at finite
temperature
\begin{equation}
\langle {\mathcal O}\rangle=\int  D U\,\TR[\exp\{-\beta {\mathsf H}\}{\mathcal O}g(U)]\, ,
\end{equation}
where $g(U)$ is the second quantized operator of the gauge
transformation represented by the matrix $U$. This expression for
fixed $U$ can be compared to the same expression but with $U$
gauge transformed
\begin{align}
{\rm Tr}[\exp\{-\beta {\mathsf H}\}{\mathcal O}g(V^\dagger UV)]
&={\rm Tr}[\exp\{-\beta {\mathsf H}\}{\mathcal O}g(V^\dagger) g(U)g(V)]\nonumber\\
&={\rm Tr}[\exp\{-\beta {\mathsf H}\}{\mathcal O}g(U)]\, .
\end{align}
The last equality here follows from the fact that both ${\mathcal O}$ and
$\exp\{\beta {\mathsf H}\}$ are gauge invariant, and thus the
operator $g(V^\dagger)$ can be commuted all the way to the left.
The only effect of the transformation is then to change the basis
over which the trace is being taken, which obviously leaves the
trace invariant.

Our variational setup is somewhat different. Expectation values
are calculated as
\begin{equation}
\int  D U {\rm Tr}[\tilde\varrho g(U) {\mathcal O}]
\end{equation}
with $\tilde\varrho$ defined in eq.~(\ref{tilderho}). This
expression is altogether gauge invariant, since the integral over
$U$ correctly projects only the contribution of gauge singlet
states. However the operator $\tilde\varrho$ is not itself
explicitly gauge invariant. For that reason the gauge
transformation operator $g(V^\dagger)$ cannot be commuted through
it, and thus
\begin{equation}
{\rm Tr}[\tilde\varrho {\mathcal O}g(V^\dagger UV)]\ne{\rm Tr}[\tilde\varrho
{\mathcal O}g(U)]
\end{equation}
even for gauge invariant operators ${\mathcal O}$. This manifests itself as
absence of local gauge invariance in the action of the effective
$\sigma$-model, eq.~(\ref{eq:action}).

Nevertheless, we stress again that since the integration over the
$SU(N)$ valued field $U$ projects out the physical Hilbert space,
its meaning in this sense is the same as that of the Polyakov
loop.

\subsection{The Calculation of the free energy}

To find the best variational density matrix we have to minimize
the free energy with respect to the variational parameters $M$ and
$H$. The Helmholtz free energy ${\sf F}$ of the density matrix
$\varrho$ is given by
\begin{equation}
    {\mathsf F} = \langle {\mathsf H} \rangle
     - {\mathsf T}  {\mathsf S}\, ,
    \end{equation}
where ${\sf H}$ is the standard Yang-Mills Hamiltonian
eq.~(\ref{ymham}), ${\sf S}$ is the entropy, and ${\sf T}$ is the
temperature.

Thus
\begin{equation}
\label{eq:freeenergy}
     {\mathsf F}
    = \frac{1}{2}\Big(
    \TR ({E}^2\varrho)
    +\TR ({B}^2\varrho)
    \Big)
    + {\mathsf T}\cdot\TR( \varrho \ln\varrho) \, .
\end{equation}
First of all we need to perform the integration over the gauge
fields, and reduce this expression to the average of a
$U$-dependent operator in the effective $\sigma$-model. In fact,
as we shall see soon, to leading order in $H$ the only non-trivial
calculation we need to perform is that of the entropy.

We will calculate the entropy up to the first non-trivial order in
$H$. As we now show, the leading term at small $H$ is $O(H\ln H)$.

Let us denote by $\varrho_{0}$ the density matrix of the pure
state with $H=0$:
\begin{equation}
    \varrho_{0} = |0\rangle\langle 0|\, .
\end{equation}
Here $|0\rangle$ does not denote necessarily the actual ground state,
but rather a projected Gaussian state with arbitrary $M$.  Now,
since the matrix elements of the density matrix can be expanded in
powers of $H$, to leading order we can write
\begin{equation}
    \varrho = \varrho_{0} + \delta\varrho\, ,
\end{equation}
where $\delta\varrho$ is $O(H)$.

Imagine that we have diagonalized $\varrho$. It will have one
large eigenvalue $\alpha_{0} = 1 -O(H)$, which corresponds to the
eigenstate
\begin{equation}
    |0'\rangle = |0\rangle +O(H)\, .
\end{equation}
All the rest of the eigenvalues $\alpha_{i}$ are at most  $O(H)$.
Then the entropy can be written as
\begin{equation}
    \label{eq:entropy1}
    {\mathsf S}= - \TR(\varrho\ln\varrho)
    = -\alpha_{0}\ln\alpha_{0} - \sum_{i=1}^{\infty}\alpha_{i}\ln\alpha_{i}\, .
\end{equation}

The second term is $O(H\ln H)$, and it is the coefficient of this
term that we will now calculate. Neglecting $O(H)$ corrections,
we can substitute $\alpha_{i}=H/M$ under the logarithm. Thus to
leading logarithmic order
\begin{equation}
    {\mathsf S} = -  \sum_{i}\alpha_{i}\ln H/M\, .
\end{equation}

Thus we have to calculate $\sum_{i}\alpha_{i}$. Let
\begin{equation}
    |0'\rangle = |0\rangle + H |x\rangle\, .
\end{equation}
Then
\begin{equation}
    \varrho = \alpha_{0} |0'\rangle\langle 0'|
    + \sum_{i=1}^{\infty}\alpha_{i} |x_i\rangle\langle x_i|
\end{equation}
with
\begin{equation}
    \langle x_{i}|0'\rangle = 0\, .
\end{equation}

Note that $\langle 0|x\rangle \neq 0$, but
\begin{equation}
    \langle 0|x\rangle + \langle x|0\rangle = 0\, ,
\end{equation}
since $|0'\rangle$ has to be normalized at $O(H)$. Also
\begin{equation}
    \langle x_{i}|0\rangle + H\langle x_{i}|x\rangle = 0\, .
\end{equation}
Thus the overlap $\langle x_{i}|0\rangle$ is $O(H)$, and we have
\begin{equation}
    \varrho = \alpha_{0} |0\rangle\langle 0|
    + H \Big( |0\rangle\langle x| + |x\rangle\langle 0|\Big)
    + \alpha_{i} |x_{i}\rangle\langle x_{i}|\, .
\end{equation}
Multiplying this by $\varrho_{0}$ we get
\begin{eqnarray}
    \varrho_{0}\cdot\varrho
    &= &\alpha_{0}\cdot\varrho_{0} +H |0\rangle\langle x|
    + H \langle 0|x \rangle |0\rangle\langle 0| \, ,\\
    \varrho\cdot\varrho_{0}
    &=& \alpha_{0}\cdot\varrho_{0} +H |x\rangle\langle 0|
    + H \langle x|0 \rangle |0\rangle\langle 0|\, .\nonumber
\end{eqnarray}
Thus,
\begin{equation}
    \varrho_{0}\cdot\varrho + \varrho\cdot\varrho_{0} - \varrho =
    \alpha_{0}\cdot\varrho_{0} - \alpha_{i} |x_{i}\rangle\langle x_{i}|\, .
\end{equation}
Multiplying again by $\varrho_{0}$, we get rid of
$|x_{i}\rangle\langle x_{i}|$ to $O(H)$
\begin{equation}
    \alpha_{0}\cdot\varrho_{0}=
    \varrho_{0}\cdot\varrho
    + \varrho_{0}\cdot\varrho\cdot\varrho_{0}
    - \varrho_{0}\cdot\varrho =
    \varrho_{0}\cdot\varrho\cdot\varrho_{0}\, .
\end{equation}
Then,
\begin{equation}
    \alpha_{0}=
    \TR(\varrho_{0}\cdot\varrho)\, .
\end{equation}
Since $\TR\varrho =1$ we have
\begin{equation}
    \sum_{i}\alpha_{i} = 1 -\alpha_{0} = \TR(\varrho_{0}(1-\varrho))
\end{equation}
which, inserted into eq.~(\ref{eq:entropy1}), gives
\begin{equation}
    {\mathsf S}= - (1-\TR(\varrho_{0}\cdot\varrho))\ln H/M \, .
\end{equation}

The derivation has been given for the normalized density matrices
$\varrho_{0}$ and $\varrho$. In terms of our Gaussian matrices we
should restore the normalization factors $Z$ and $Z_0$, so that
finally we have
\begin{equation}
    {\mathsf S}
    =\Big(
    \frac{\TR(\varrho_{0}\cdot\varrho)}{\TR\varrho_{0}\cdot\TR\varrho}-1
    \Big) \ln H/M \, .
\end{equation}
It is easy to check that to $O(H)$
\begin{equation}
    \TR(\varrho_{0}\cdot\varrho) = (\TR\varrho_{0})^{2}\,
\end{equation}
and
\begin{equation}
    {\mathsf S}
    =\Big(
    \frac{\TR\varrho_{0}}{\TR\varrho}-1
    \Big) \ln H/M \, .
\end{equation}

From eq.~(\ref{eq:trvarrho}), it is clear that
\begin{equation}
    \TR\varrho =
    \bigg[
    1+ H\Big(
    \frac{1}{4\pi^2}M^2\tr S
    -\frac{1}{4}\lambda S \lambda
    \Big)
    \bigg] \cdot \TR\varrho_{0}\, .
 \end{equation}
Using eqs.~(\ref{eq:lr},\ref{eq:trs}) we finally get
\begin{multline}
    {\mathsf S}
    = -\biggl[\Big\langle
    \frac{1}{8 g^{2}}
    \tr (U^\dagger \partial U - \partial U^\dagger U)
    (\partial U U^\dagger - U\partial U^\dagger)\\
    + \frac{1}{4\pi^2}M^2 (\tr U^{\dagger}\tr U -1)\Big\rangle_U
    \biggr]
    H\ln H/M\, .
    \label{entr}
\end{multline}

To leading order in $H$, the averaging over $U$ in this expression
has to be performed with the $\sigma$-model action with $H=0$.

The expression of eq.~(\ref{entr}) has the following striking property.
For $M<M_c$ it vanishes identically. The reason is very simple.
The first term in eq.~(\ref{entr}) is the product of the left
handed $SU(N)$ current and the right handed $SU(N)$ current in the
$\sigma$-model. Thus it transforms as an adjoint representation
under each one of the $SU(N)$ factors of the $SU_L(N)\otimes
SU_R(N)$ transformation. The same is also true for the second term
in eq.~(\ref{entr}). The $\sigma$-model action at $H=0$
is itself obviously  invariant under the whole $SU_L(N)\otimes SU_R(N)$
group. Now, at $M<M_c$, the symmetry group is not spontaneously
broken, and thus any operator which is not a scalar has a
vanishing expectation value. It follows immediately that the
entropy has an $O(H\ln H/M)$ contribution only for $M>M_c$, when
the $SU_L(N)\otimes SU_R(N)$ group is spontaneously broken down to
$SU_V(N)$.

This observation makes our task considerably simpler. Since for
$M<M_c$ the entropy is zero, we do not have to consider at all the
disordered phase of the effective $\sigma$-model. In this
disordered phase the free energy coincides with energy, and thus
the calculation is identical to the calculation at zero
temperature presented in sec.~(\ref{sec:ym}).

 Thus we only need to consider the effective
$\sigma$-model in the ordered phase. As at $\mathsf{T}=0$ we
perform the calculations in the ordered phase to leading order in
$\alpha_s$. Since there are no $O(H\ln H/M)$ corrections to energy
at this order, the result for the energy in the disordered phase
is identical to the result at zero temperature,
eq.~(\ref{largeM}). Thus our expression for the free energy in the
ordered phase of the $\sigma$-model is
\begin{multline}
{\mathsf F}={N^2\over 120\pi^2}M^4+{\mathsf T}\biggl(\Big\langle
    \frac{1}{8 g^{2}}
    \tr (U^\dagger \partial U - \partial U^\dagger U)
    (\partial U U^\dagger - U\partial U^\dagger)\\
    + \frac{1}{4\pi^2}M^2 (\tr U^{\dagger}\tr U -1)\Big\rangle_U
    \biggr)
    H\ln H/M\, .
\end{multline}

We now average over $U$ in the leading order perturbation theory.

\subsection{The $\sigma$-model perturbation theory}

For the purpose of the perturbative $\sigma$-model calculation we
parameterize the $U$ matrices as

\begin{equation}
    U =    \exp\Big\{\frac{i}{2} g \phi^{a}\tau^{a}\Big\}\, .
    \label{pertu}
\end{equation}

Although we only need the leading order, it is instructive to
check that the order $g^{2}$ term in the expansion is indeed
small. To this order we have
\begin{equation}
    U \simeq
    \Big(1 + \frac{i}{2} g \phi^{a}\tau^{a}
    - \frac{1}{8} g^2 \phi^{a}\phi^{b}\tau^{a}\tau^{b}
    - \frac{i}{48} g^3 \phi^{a}\phi^{b}\phi^{c}\tau^{a}\tau^{b}\tau^{c}
    \Big)\, .
\end{equation}
So that the $\sigma$-model action becomes
\begin{eqnarray}
    {\mathcal S} &=& {M\over 2g^2}\tr(\partial U \partial U^{\dagger})
    \nonumber\\
    &=& \frac{M}{4} \partial\phi\partial\phi
    + \frac{M}{192} g^{2} (\partial\phi^{a})(\partial\phi^{c})
    \phi^{b}\phi^{d}
   \tr \Big[
   \tau^{a}\tau^{b}\tau^{c}\tau^{d}
    -\tau^{a}\tau^{c}\tau^{b}\tau^{d}
     \Big]\, .
\label{pertact}
\end{eqnarray}
The propagator of the phase field $\phi$ is thus
\begin{equation}
    \langle \phi^{a}\phi^{b} \rangle =
    \frac{2}{M k^{2}} \delta^{ab}\, .
\end{equation}
To get the idea of the quality of this perturbative expansion we
can calculate for example $\langle {\mathcal S}\rangle$. In this calculation
one has to take into account the fact that the measure in the path
integral over the phase $\phi^a$ is not the simple $ D \phi$,
but rather the group invariant $U(N)$ measure $\mu$ . To first
order in $g^2$ it is
\begin{equation}
\mu= D \phi^a \exp\Big\{{M^3N\over 144\pi^2} g^{2} \int d^3x \phi^2(x)\Big\}\, .
\label{measure}
\end{equation}
Taking this into account we find that $\langle {\mathcal S}\rangle$ gets no
correction of order $g^2$. We thus feel confident that the use of
the perturbation theory in the ordered phase of the $\sigma$-model
is an admissible approximation.
 In the following we will only keep leading order expressions.

Calculating to leading order the entropy eq.~(\ref{entr}) and
keeping only the $O(N^2)$ terms we find
\begin{equation}
    \langle {\mathsf S} \rangle
    =-\frac{N^{2}}{6\pi^{2}}M^{2} H\ln {H\over M}\, .
\end{equation}

Introducing the dimensionless quantity
\begin{equation}
    h=\frac{H}{M}
\end{equation}
we can write the expression for the free energy as
\begin{eqnarray}
     {\mathsf F}  &=&
    \langle {\mathsf H} \rangle
    -{\mathsf T}\langle {\mathsf S} \rangle \nonumber \\
    &=& \frac{N^{2}}{120\pi^{2}}
    M^{4}
    + {\mathsf T}\frac{N^{2}}{6\pi^{2}}M^{3}  h\ln  h\, .
\end{eqnarray}
We now have to minimize this expression with respect to $h$ and
$M$. It is convenient to first perform the minimization with
respect to $h$ at fixed $M$. This obviously gives
\begin{equation}
    \frac{\partial {\mathsf F}} {\partial h}=0
    \rightarrow {h}=\frac{1}{e}\, .
\end{equation}
Thus as a function of $M$ only, the free energy becomes
\begin{equation}
     {\mathsf F}
    = \frac{N^{2}}{120\pi^{2}}
    M^{4}
    - \frac{\mathsf T}{e}\frac{N^{2}}{6\pi^{2}}M^{3}\, .
\end{equation}
Now minimizing with respect to $M$ we find
\begin{equation}
    \frac{\partial {\mathsf F}} {\partial M}=0
    \rightarrow M = \frac{15 {\mathsf T}}{e}\, .
\end{equation}

Thus for $M\ge M_c$ the free energy of the best variational
density matrix as a function of temperature is
\begin{equation}
    {\mathsf F}_{M\ge M_c}
    = -\frac{N^{2}}{360\pi^{2}}
    \Big(\frac{15 {\mathsf T}}{e}\Big)^{4}\, .
    \label{freeen}
\end{equation}

We now have to compare this value with the free energy for $M\le
M_c$. As we have discussed above, this is given by the expectation
value of the Hamiltonian alone, and is minimized at $M=M_c$. Its
value is
\begin{equation}
{\mathsf F}_{M\le M_c}=-{N^2\over 30\pi^2}M_c^4\, .
\label{freeen1}
\end{equation}
Comparing the two expressions we find
\begin{equation}
\mathsf{T_{c}}= \frac{12^{\frac{1}{4}} e}{15} M_c\, .
\end{equation}
Using the value of $M_c$ from eq.~(\ref{mc}) we have
\begin{equation}
\mathsf{T_{c}}= 450\, \mathrm{Mev}\, .
\end{equation}

For $\mathsf{T}\le \mathsf{T_{c}}$ the free energy is minimized in
the variational state with $M=M_c$. In our approximation this
state is the same as at zero temperature. Its entropy vanishes,
and the effective $\sigma$-model is in the disordered phase. The
Polyakov loop vanishes, $\langle U\rangle=0$ and according to the
standard wisdom this is a confining state.

For $\mathsf{T}\ge \mathsf{T_{c}}$ the best variational state is very different. The
entropy of this state is non-zero,
\begin{equation}
 {\mathsf S}={N^2\over 6\pi^2 e}\Big({15 {\mathsf T}\over e}\Big)^3\, .
\end{equation}
The Polyakov loop is non-zero $\langle U\rangle\ne 0$ and thus the high
temperature density matrix describes a deconfined phase.

Finally, we note that in the deconfined phase our best variational
density matrix has a non-vanishing ``electric screening" or ``Debye"
mass. The Debye mass is conveniently defined as the ``mass" of the
phase of the Polyakov loop. This mass is non-vanishing in our
calculation for the following reason. As long as $H=0$, the
effective $\sigma$-model action has a global $SU_L(N)\otimes
SU_R(N)$ symmetry. Thus in the ordered phase of the $\sigma$ model
the phases $\phi^a$ are massless. However, as discussed above, the
terms of order $H$ in eq.~(\ref{eq:action}) break this symmetry
explicitly down to the diagonal $SU_V(N)$. As a result the would
be ``Goldstone" phases $\phi^a$ acquire mass. To calculate this
mass it is convenient first to note that to $O(g^2)$
\begin{multline}
\tr(U^\dagger \partial U - \partial U^\dagger U)
    (\partial U U^\dagger - U\partial U^\dagger)\\
    =-4\, \tr (\partial U^\dagger
    \partial U)
    -{g^4 \over
    4}  \phi^a\phi^c\partial\phi^b\partial\phi^d
    \tr\big(\tau^a\tau^b\tau^c\tau^d-\tau^a\tau^c\tau^b\tau^d\big)\, .
    \label{interact}
\end{multline}
The contribution of the $SU_L(N)\otimes SU_R(N)$ term to the mass
cancels against the contribution of the measure
eq.~(\ref{measure}).
 Using
eqs.(\ref{eq:action}, \ref{interact}) we then find to $O(g^2)$ and
to leading order in $H$
\begin{equation}
M_D^2={4\over 3\pi}\alpha_s(M)NMH \, .
\end{equation}

As a function of temperature we have
\begin{equation}
M_D^2=\alpha_s\Big({15\over e}T\Big)N{300\over \pi e^3}{\mathsf T}^2\, .
\label{md}
\end{equation}

Let us summarize the results of our analysis of the deconfinement transition.
We find the phase transition at a temperature of about
$\mathsf{T_{c}}\simeq 450\, {\rm Mev}$. The transition is strongly
first order at large $N$. The latent heat is $\Delta E={N^2\over
90\pi^2 }\big ({15 {\mathsf T}\over e}\big)^4$. Below the
transition the entropy is zero, the best variational state is the
same as at zero temperature, and the average value of the Polyakov
loop is zero. Above the transition, the entropy is non-zero and
proportional to the number of ``coloured" degrees of freedom,
${\mathsf S}\propto N^2$. The average value of the Polyakov loop
is non-zero and the phase is deconfined.

It is quite interesting that at high temperature our formulae
numerically are quite close to the predictions of free gluon
plasma. In particular, our value for the free energy,
eq.~(\ref{freeen}), should be compared to the free gluon plasma
expression
\begin{equation}
{\mathsf F}_{free}= -\frac{N^2\pi^2}{45} {\mathsf T}^4\, .
\end{equation}
The ratio between the two is
\begin{equation}
\frac{{\mathsf F}_{free}}{{\mathsf F}_{var}}\simeq 0.85 \,
.\label{ratiof}
\end{equation}
The ratio of the entropies is the same.

Interestingly we get the same ratio  comparing our value for the
Debye mass eq.(\ref{md}) with the leading order perturbative one,
$M_{pert}^2={4\pi\over 3}\alpha_sNT^2$,
\begin{equation}
{M_{pert}^2\over M_D^2}\simeq 0.85 \, .\label{ratio1}
\end{equation}

The pressure approaches its asymptotic value according to the
simple formula
\begin{equation}
\frac{{\mathsf P}({\mathsf T})}{{\mathsf P}_{\rm asympt}}=1-\frac
{\mathsf{T_{c}}^4}{{\mathsf T}^4}\, .
\end{equation}
Here the asymptotic value of the pressure ${\mathsf P}_{\rm
asympt}$ is given by eq.~(\ref{freeen}). The pressure ${\mathsf
P}({\mathsf T})$ is given by the difference between
eq.~(\ref{freeen}) and the value of the free energy at zero
temperature, which coincides with expression eq.~(\ref{freeen1}).

One has to take the comparison eqs.~(\ref{ratiof},\ref{ratio1})
with a grain of salt. As explained above, our calculations were
performed assuming small $H$. A priori we expect that this
restriction should confine us to not too large temperatures. On
the other hand the minimization of the free energy resulted in the
value $H/M=1/e$ independently of temperature. Thus, we feel that
the comparison eq.~(\ref{ratiof}) may be meaningful.

The main features of these results are indeed what we expect from
the deconfinement phase transition on general grounds. It is nice
that a simple minded calculation such as this
 does qualitatively so well in such a complicated problem.
It therefore appears that the projection of the trial density
matrix on the gauge invariant Hilbert space is, just like at zero
temperature, the crucial feature that dictates most if not all the
important aspects of the low energy and low temperature physics.
In the context of the present calculation the most important
effect of the gauge projection is obviously vanishing of the
entropy in the low temperature phase. We stress that this feature
was not at all built into our initial ansatz, but followed
naturally and unavoidably in the disordered phase of the effective
$\sigma$-model.

 Quantitatively, this calculation of course should be taken for what
 it is --- an approximate implementation of the variational
 principle. As with any variational calculation, the range of validity of
 this calculation is
 not sharply defined. Even within the variational framework we had to
 resort to additional approximations.
 The most severe simplifications that we had to impose
 are the perturbation theory in the ordered phase of the $\sigma$-model
 and the assumption of smallness of $H$.
 The projection over the gauge group, which as we saw
 is so physically important, is what makes the calculational task
 difficult and forces us to make these approximations.

 The assumptions of smallness of $g$ and of smallness of $H$ affect
different aspects of our result. In particular, in the leading
order of the perturbation theory the expectation value of the
Polyakov loop $U$ is equal to unity. The actual value of $U$ on
the ordered side of the transition according to \cite{Kogut:1982ez} is
close to one half. Thus our perturbative calculation is rather
more reliable somewhat further away from the transition. The
closer to the transition we get, the more important higher order
corrections in $g^2$ become. Thus to properly describe the
transition region of QCD we need to improve our calculational
method in the vicinity of the transition in the $\sigma$-model.
 In line with this we expect that the
estimate for the critical temperature we obtained here is somewhat
higher than we would get, had we treated the $\sigma$-model more
accurately in the transition region. This is consistent with the
fact that our result for $\mathsf{T_{c}}$ is by about $50\%$
higher than the lattice value of $270\, {\rm Mev}$.

The smallness of $h$ is quite important in a different way. The
value of $h=1/e$ that we obtain is in fact a reasonably small
number, so omitting the corrections in powers of $h$ is fairly
safe. On the other hand, the terms linear in $h$ but not enhanced
by $\ln h$, which we have ignored in the present calculation, have
to be accounted for more carefully. With the value of $h$ that we
obtain, these terms are not suppressed in any obvious way.

The obvious stumbling block to any improvement along these lines is the calculation of the entropy
${\mathsf S}=-\tr\varrho\ln\varrho$. However, if one opts for restricting, as before, the analysis
to leading order in $g^2$ the entropy and, therefore, the free energy can be, without any additional
approximations, calculated to all orders in $h$ \cite{Gripaios:2003ce}. This improved analysis is
carried out in the following.

\subsection{All-order in $h$ analysis}

Let us ask ourselves what would happen if we did not restrict $H$
to be small, and more generally did not restrict the functional
forms of $G(k)$ and $H(k)$ in our variational ansatz. We could
still carry on our calculation for a while. Namely we would be
able to integrate over the vector potentials in all averages, and
would reduce the calculation to a consideration of some non-linear
$\sigma$-model of the $U$-field. This $\sigma$-model quite
generally will have a symmetry breaking phase transition as the
variational functions $G(k)$ and $H(k)$ are varied. Since at this
transition the Polyakov loop $U$ changes its behaviour, the
disordered phase of the $\sigma$-model corresponds to the
confining phase of the Yang Mills theory, while the ordered phase
of the $\sigma$-model represents the deconfined  phase. Thus, in
order to study deconfinement in the $SU(N)$ Yang Mills theory, we
should analyze the physics of each $\sigma$-model phase as accurately
as possible and calculate the transition scale $M_c$ (or rather
$G_c(k)$). We then calculate the free energy of the $\sigma$-model
in each phase at temperature $T$ and extract the minimal free
energy. The deconfinement transition occurs at the temperature for
which the free energies calculated in the ordered and disordered
phases of the sigma model coincide.

In practice in the disordered phase no progress seems possible
without restricting the arbitrary kernels and we adopt the forms eqs.~(\ref{G},\ref{H}). The resulting
minimal free energy is thus independent of the temperature and is
given by eq.~(\ref{freeen1}).

On the other hand, in the ordered phase we can relax the
restriction on $H$ and $G$ if, as before we work in the leading
order in perturbation theory. In this case  minimization with
respect to arbitrary kernels $G^{-1}(k)$ and $H(k)$ is possible.
We now describe this calculation following \cite{Gripaios:2003ce}.

In this approximation, for the $U$ matrices we use the
parameterization eq.(\ref{pertu}). Hence at leading order one can
take
\begin{align}
U &\simeq 1\, , \nonumber \\
\partial_i U &\simeq  ig \partial_i \phi^a \frac{\tau^a}{2}.
\end{align}
Thus, the gauge transformations eq.~(\ref{gtqcd}) reduce to
\begin{gather}
A^{a}_{i} \rightarrow A^{a}_{i} - \partial_i \phi^a
\end{gather}
and the Hamiltonian eq.~(\ref{ymham}) reduces to
\begin{gather}
\mathsf{H} = \frac{1}{2} \left[ E^{a2}_{i} + (\epsilon_{ijk} \partial_j A^{a}_{k})^2 \right]\, .
\end{gather}
These last two equations describe the theory $U(1)^{N^2-1}$: in
the leading order of the $\sigma$ - model perturbation theory, the
$SU(N)$ Yang--Mills theory reduces to the $U(1)^{N^2-1}$ free
theory. The density matrix eq.~(\ref{eq:ansatz}) becomes Gaussian
again, because the gauge transformations are linear. One has
\begin{multline} \label{gans}
\varrho [A,A^{'}] = \int D\phi \; \exp\bigg\{-\frac{1}{2} \Big[ A
G^{-1} A + (A' - \partial \phi) G^{-1} (A' - \partial \phi)
\\- 2 A H (A' - \partial \phi) \Big] \bigg\}\, .
\end{multline}

Now the theory of $N^2-1$ $U(1)$ free fields in $3+1$ dimensions
is completely tractable. Both the energy and the entropy can be
calculated explicitly \cite{Gripaios:2002xb}. The free energy in
terms of the arbitrary kernels $G^{-1}$ and $H$ is
\begin{multline}
{\mathsf F} = \frac{N^2-1}{2} \int \frac{d^3p}{(2\pi)^3}
\Bigg[  G^{-1}(1 + GH) + p^2 G (1 - GH)^{-1} \\
- 4{\mathsf T}  \left(  \ln \bigg[ \frac{GH}{\xi}\bigg]
 - \ln \bigg[ \frac{\eta}{GH}\bigg]
 \cdot \frac{\eta}{\xi}  \Bigg) \right]\, ,
\end{multline}
where $\eta=1- (1-(GH)^2)^{1/2}$ and $\xi=(1-(GH)^2)^{1/2}-(1-GH)$.

It is  minimized by
\begin{align} \label{mkers} G^{-1} &= p
\left( \frac{1+ e^{-\frac{2p}{\mathsf T}}}{1 - e^{-\frac{2p}{\mathsf T}}}\right), \nonumber \\
H &= 2p
 \left( \frac{e^{-\frac{p}{\mathsf T}}}{1 - e^{-\frac{2p}{\mathsf T}}}\right)
\end{align}
and the minimal value of the free energy at temperature $\mathsf T$ is
\begin{align} \label{minpf}
{\mathsf F} &= \frac{N^2-1}{\pi^2} \int_{0}^{\infty} p^2dp \; \left[ \frac{p}{2} + {\mathsf T}
\ln (1 - e^{-p/{\mathsf T}})\right] \nonumber \\
  &= - \frac{(N^2-1){\mathsf T}^4}{3 \pi^2} \int_{0}^{\infty} dx \frac{x^3}{e^{x} -1} \nonumber \\
  &= - \frac{\pi^2(N^2-1){\mathsf T}^4}{45}\, ,
\end{align}
where, just like in eq.~(\ref{freeen1}), zero-point term has been
discarded. This is, of course, just the free energy of a free photon
gas.

Thus the free energy of $SU(N)$ Yang Mills theory is minimized
with $M=M_c$ in the disordered phase of the $\sigma$- model for
temperatures below ${\mathsf T}_c$, which is obtained by equating
the free energies eqs.~(\ref{freeen1}) and (\ref{minpf}).
\begin{gather}
- \frac{N^2 M_{c}^{4}}{30 \pi^2} =  - \frac{\pi^2 N^2 {\mathsf
T}_{c}^{4}}{45}\, ,
\end{gather}
which yields
\begin{gather}
{\mathsf T}_c = \left( \frac{3}{2}\right)^{1/4} \frac{M_c}{\pi} \simeq 470\, \mathrm{MeV}\, .
\end{gather}

The all order in $h$ improvement discussed here allows us to take
more seriously our results at high temperature. At high
temperatures the kernel, which corresponds to the Boltzmann
factor,  is of order unity and thus our original assumption of
smallness of $H$. This is indeed obvious from eq.~(\ref{mkers}). As
a result we now reproduce the expected asymptotic free gluon
plasma result for the free energy.

On the other hand this improvement affected very little our
previous results in the transition region. The transition
temperature is shifted  only by about $5\%$. The same is true for
the the value of the parameter $H$ at low momentum. As before, we
find that the deconfinement phase transition is strongly first
order with latent heat $\Delta E=\frac{4\pi^2 N^2}{45} {\mathsf
T}_{c}^{4}$.

Although the actual value of the transition temperature is
considerably larger than the lattice estimate, as explained
earlier it makes more sense to look at dimensionless quantities.
In particular, if we identify $2M_c$ with the mass of the lightest
glueball (see however  \cite{Gripaios:2002bu} and the discussion
in the previous section), we find
\begin{gather} \label{ratio}
\frac{{\mathsf T}_c}{2M_c} = \frac{1}{2\pi}\left(\frac{3}{2}\right)^{1/4} \simeq 0.18
\end{gather}
This is in excellent agreement with the lattice estimate for
$SU(3)$ pure gauge theory \cite{Teper:1998kw}. We should however
caution that given the uncertainties in our calculation this
agreement may well be fortuitous.

\section{Conclusions}

We have tried here to critically review the application of the
variational principle to Quantum Field Theories with gauge
invariance, with the main focus  on the approach developed by Ian
Kogan and collaborators
\cite{Kogan:1995wf,Kogan:1995vb,Brown:1997gm,Brown:1998cp,Kogan:2002yr}.

Although it is too early to decide whether this approach can be a
useful calculational scheme for strongly interacting gauge
theories, we can draw encouragement from its performance in the
non-trivial toy models. In particular, in compact QED in $2+1$
dimensions, we have been able to reproduce all known non-trivial
characteristics of the non-perturbative vacuum state: dynamical
mass generation, confining potential between external charges and
area law behaviour of the spatial Wilson loop with parameterically
correct values for the string tension and mass. Although this is
the only example that we have covered extensively in this review,
the method has also been applied to other lower dimensional
systems, and it works very well in all cases. Thus the deconfining
phase transition in 2+1 compact QED at finite temperature is
described correctly \cite{Gripaios:2002bu}. In the (exactly
solvable) Schwinger model the variational approach reproduces the
exact ground state wave functional\cite{schwinger}. In the compact
2+1 QED with Chern-Simons term \cite{chernsimons} it predicts a
Kosterlitz-Thouless phase transition in the value of the Chern-Simons parameter, in agreement with earlier analysis
\cite{ianvortices}.

In 3+1 dimensional gluodynamics, this variational method gives
results which on the qualitative level at least, conform with our
intuition about the structure of the ground state, both at zero
and finite temperature. We find dynamical mass generation,
corresponding to an acceptable value of the gluon condensate. At
finite temperature we find a first order phase transition which
corresponds to the Polyakov loop acquiring a non-zero average.
Although we have not calculated the string tension directly, the
behaviour of the Polyakov loop is very much indicative that this is
indeed the deconfining phase transition. The value of the critical
temperature (in units of glueball mass) we find is in good
agreement with lattice results. We also found that in the low
temperature phase the entropy remains zero all the way up to the
transition temperature. This is a rather striking result, which
has not been built into our variational ansatz, but rather emerged
as the result of the dynamical calculation.

An important lesson we learned from the lower dimensional models
is that the projection of the Gaussian trial state onto the gauge
invariant Hilbert subspace dictates most, if not all, of the
important aspects of the non-perturbative physics. It was
absolutely essential to perform the projection non-perturbatively,
fully taking into account the contribution of the overlap between
gauge rotated Gaussians into the variational energy prior to
minimization.

The same conclusion carries over to the pure Yang Mills theory. We
have seen that from the point of view of the effective
$\sigma$-model the energy is minimized in the disordered phase. In
other words, the low momentum fluctuations of the field $U$ are
large, unlike in the perturbative regime, where $U$ is close to a
unit matrix. From the point of view of the trial wave functional,
this means that the off-diagonal contributions, coming from the
Gaussian WF gauge rotated by a slowly varying gauge
transformation, are large. It is these ``off diagonal" contributions
to the energy that lowered the energy of the best trial state
below the perturbative value. In the low temperature phase the
vanishing of the entropy was also a direct consequence of the
effective $\sigma$-model being in the disordered phase, and thus
of the non-perturbative nature of the gauge projection. The
accounting for these off diagonal terms non-perturbatively is the
main distinction between this approach and other attempts
\cite{Kerman:1989kb,Heinemann:1999ja,Schroeder:2003qw,Preparata:1986cq,
Consoli:1985sx,Preparata:1986bk} to implement the variational
principle in gauge theories.

Many outstanding questions remain. Is the best variational state
confining? How do we calculate the interaction potential between
external sources? How do we understand better the relation between
the variational parameter and the glueball masses? Can we extend
the Ansatz to include (massless) fermions?

Both to be confident in our results and to be able to approach
these questions we need first and foremost to have a better way of
treating analytically the effective non-linear $\sigma$-model. The
use of the mean field approximation in the effective
$\sigma$-model was the main source of uncertainties in our
calculations both at zero and finite temperature.  We believe that
it should be possible to treat the $\sigma$-model in a better way,
perhaps along the lines of a continuum version of
\cite{Kogut:1982ez}. Such an improvement is crucial to clarify
whether the qualitatively appealing results that we have described
here are a kind of fluke due to an interplay of two bad
approximations (variational and mean field) or are genuine
predictions of a useful, workable variational approach. Personally
we do believe that these results are genuine and that there is
enough scope for further development of the approach which
warrants continuing active investigations.

\section*{Acknowledgments}
This review is dedicated to the memory of Ian Kogan. Ian was an
extraordinary physicist with an almost unimaginable breadth of
interests. The non-perturbative domain of QCD was but one of the many
problems that interested him. As in all areas that he worked in,
Ian left a lasting mark in this subject. During the last 10 years
of his life, Ian periodically returned to the variational approach
described in this review, always with new ideas of what to do next
and how to do it better. Ian's enthusiasm and bubbling energy was
the driving force that caused the initial embryonic idea to
develop into a solid calculational approach which has already
produced many interesting results, and will hopefully keep on
developing and improving. Ian's untimely death is a severe blow to
all of his many friends. We miss him...

We want to thank our collaborators Will Brown, Juan Pedro
Garrahan, Ben Gripaios and Ben Svetitsky for an enjoyable
collaboration spanning several years. 

One of us (J.G.M.) wishes to express his gratitude to the NA60 collaboration, especially to Carlos Louren\c co and Jo\~ao Seixas, for their hospitality.
The work of J.G.M. was supported by the Funda\c c\~ao para a Ci\^encia e a Tecnologia (Portugal) under contract SFRH/BPD/12112/2003.



\begin{thebibliography}{10}


\bibitem{lattice}M. Creutz, Yang-Mills fields and the
lattice, [hep-lat/0406007]

\bibitem{effective} H. Leutwyler, in M. Shifman,(ed.): At the
frontier of particle physics, vol. 1 271; [hep-ph/0008124]



\bibitem{monopolevortex} I. Kogan and A. Kovner, in M. Shifman (ed): At the frontier of particle physics, vol. 4, 2335; [hep-th/0205026];  

\bibitem{mongreen} J. Greensite, Progr. Part. Nucl. Phys.51:1 (2003); [hep-lat/0209138];



\bibitem{susy} See  A. Armoni, M. Shifman and G. Veneziano in this volume, [hep-th/0403071]

\bibitem{vangerooge} R. Feynman, in Wangerooge 1987, Proceedings,
"Variational calculations in quantum field theory", L. Polley and
D. Pottinger, eds, World Scientific (Singapore), 1988.



\bibitem{Dirac:pj}
P.~A.~M.~Dirac,
Can.\ J.\ Math.\  {\bf 2}, 129 (1950).

\bibitem{Dirac:1958jc}
P.~A.~M.~Dirac,
Phys.\ Rev.\  {\bf 114}, 924 (1959).




\bibitem{bcs} See for example J. Negele and H. Orland, "Quantum many
particle systems",  Adddison - Wesley, 1987.

\bibitem{moshe} M. Moshe and J. Zinn-Justin; \newblock Phys.Rept.{\bf 385} 69-228
(2003) [hep-th/0306133]


\bibitem{vort} A. Kovner, in M. Shifman, (Ed.)
 At the Frontier of Particle Physics, (World Scientific, Singapore, 2001) Vol. 3 1777; [hep-ph/0009138].





\bibitem{B94}
{\em Selected Papers, with Commentary, of Tony Royle Hilton Skyrme},
ed. G. Brown (World Scientific, Singapore, 1994).



\bibitem{D95} For review of the instanton physics of QCD see
D. Dyakonov,
``Chiral Symmetry Breaking by Instantons'', lectures at
the Enrico Fermi School in Physics, 1995, hep-ph/9602375.





\bibitem{Kogan:1995wf}
I.~I. Kogan and A.~Kovner,
\newblock Phys. Rev. {\bf D52}, 3719 (1995), [hep-th/9408081].

\bibitem{Shifman:1979bx}
M.~A. Shifman, A.~I. Vainshtein and V.~I. Zakharov,
\newblock Nucl. Phys. {\bf B147}, 385 (1979).

\bibitem{Nojiri:1984nu}
S.~Nojiri,
\newblock Z. Phys. {\bf C22}, 245 (1984).

\bibitem{Rosenstein:1986ku}
B.~Rosenstein and A.~Kovner,
\newblock Phys. Lett. {\bf B177}, 71 (1986).

\bibitem{Gribov:1978wm}
V.~N. Gribov,
\newblock Nucl. Phys. {\bf B139}, 1 (1978).

\bibitem{Schutte:1985sd}
D.~Schutte,
\newblock Phys. Rev. {\bf D31}, 810 (1985).

\bibitem{Cutkosky:1987nh}
R.~E. Cutkosky,
\newblock Phys. Lett. {\bf B194}, 91 (1987).

\bibitem{Cutkosky:1988yi}
R.~E. Cutkosky and K.~C. Wang,
\newblock Phys. Rev. {\bf D37}, 3024 (1988).

\bibitem{Greensite:1979yn}
J.~P. Greensite,
\newblock Nucl. Phys. {\bf B158}, 469 (1979).

\bibitem{Consoli:1985sx}
M.~Consoli and G.~Preparata,
\newblock Phys. Lett. {\bf B154}, 411 (1985).

\bibitem{Preparata:1986bk}
G.~Preparata,
\newblock Nuovo Cim. {\bf A96}, 366 (1986).

\bibitem{Preparata:1986cq}
G.~Preparata,
\newblock Nuovo Cim. {\bf A96}, 394 (1986).

\bibitem{Kerman:1989kb}
A.~K. Kerman and D.~Vautherin,
\newblock Annals Phys. {\bf 192}, 408 (1989).

\bibitem{Heinemann:1999ja}
C.~Heinemann, E.~Iancu, C.~Martin and D.~Vautherin,
\newblock Phys. Rev. {\bf D61}, 116008 (2000), [hep-ph/9911515].

\bibitem{Schroeder:2003qw}
O.~Schroeder and H.~Reinhardt,
\newblock Ann. Phys. {\bf 307}, 452 (2003), [hep-ph/0306244].

\bibitem{Kogan:1995vb}
I.~I. Kogan and A.~Kovner,
\newblock Phys. Rev. {\bf D51}, 1948 (1995), [hep-th/9410067].

\bibitem{Coleman:1975bu} S.~R. Coleman,
\newblock Phys. Rev. {\bf D11}, 2088 (1975).

\bibitem{Polyakov:1987ez}
A.~M. Polyakov, Gauge fields and strings,
\newblock Chur, Switzerland: Harwood (1987) 301 P. (Contemporary Concepts in  Physics, 3).

\bibitem{Samuel:1978vy}
S.~Samuel,
\newblock Phys. Rev. {\bf D18}, 1916 (1978).

\bibitem{Amit:1980ab}
D.~J. Amit, Y.~Y. Goldschmidt and G.~Grinstein,
\newblock J. Phys. {\bf A13}, 585 (1980).

\bibitem{Boyanovsky:1989ge}
D.~Boyanovsky,
\newblock J. Phys. {\bf A22}, 2601 (1989).

\bibitem{Polyakov:1975rs}
A.~M. Polyakov,
\newblock Phys. Lett. {\bf B59}, 82 (1975).

\bibitem{Polyakov:1977fu}
A.~M. Polyakov,
\newblock Nucl. Phys. {\bf B120}, 429 (1977).

\bibitem{Diakonov:1998ir}
D.~Diakonov,
\newblock Surveys High Energ. Phys. {\bf 14}, 29 (1999), [hep-th/9805137].

\bibitem{Zarembo:1998ms}
K.~Zarembo,
\newblock Phys. Lett. {\bf B421}, 325 (1998), [hep-th/9710235].

\bibitem{Zarembo:1998xq}
K.~Zarembo,
\newblock Mod. Phys. Lett. {\bf A13}, 2317 (1998), [hep-th/9806150].

\bibitem{Kovner:1998eg}
A.~Kovner and B.~Svetitsky,
\newblock Phys. Rev. {\bf D60}, 105032 (1999), [hep-lat/9811015].

\bibitem{Brown:1997gm}
W.~E. Brown and I.~I. Kogan,
\newblock Int. J. Mod. Phys. {\bf A14}, 799 (1999), [hep-th/9705136].

\bibitem{Gripaios:2002bu}
B.~M. Gripaios,
\newblock Int. J. Mod. Phys. {\bf A18}, 85 (2003), [hep-ph/0204310].

\bibitem{Pisarski:1984ms}
R.~D. Pisarski and F.~Wilczek,
\newblock Phys. Rev. {\bf D29}, 338 (1984).

\bibitem{Brown:1990ev}
F.~R. Brown {\em et~al.},
\newblock Phys. Rev. Lett. {\bf 65}, 2491 (1990).

\bibitem{Brown:1990by}
F.~R. Brown {\em et~al.},
\newblock Phys. Lett. {\bf B251}, 181 (1990).

\bibitem{Brezin:1973jc}
E.~Brezin, J.~C. Le~Guillou and J.~Zinn-Justin,
\newblock Phys. Rev. {\bf D8}, 2418 (1973).

\bibitem{Atiyah:1989dq}
M.~F. Atiyah and N.~S. Manton,
\newblock Phys. Lett. {\bf B222}, 438 (1989).

\bibitem{Kogut:1979wt}
J.~B. Kogut,
\newblock Rev. Mod. Phys. {\bf 51}, 659 (1979).

\bibitem{Brown:1998cp}
W.~Brown, J.~P. Garrahan, I.~I. Kogan and A.~Kovner,
\newblock Phys. Rev. {\bf D59}, 034015 (1999), [hep-ph/9808216].

\bibitem{Shuryak:1982ff}
E.~V. Shuryak,
\newblock Nucl. Phys. {\bf B203}, 93 (1982).

\bibitem{Shuryak:1982dp}
E.~V. Shuryak,
\newblock Nucl. Phys. {\bf B203}, 116 (1982).

\bibitem{Diakonov:1984hh}
D.~Diakonov and V.~Y. Petrov,
\newblock Nucl. Phys. {\bf B245}, 259 (1984).

\bibitem{Shuryak:1989fy}
E.~V. Shuryak,
\newblock Nucl. Phys. {\bf B328}, 102 (1989).

\bibitem{Shuryak:1990cx}
E.~V. Shuryak and J.~J.~M. Verbaarschot,
\newblock Nucl. Phys. {\bf B341}, 1 (1990).

\bibitem{Diakonov:1989fc}
D.~Diakonov and V.~Y. Petrov,
\newblock Phys. Lett. {\bf B224}, 131 (1989).

\bibitem{Polyakov:1988md}
A.~M. Polyakov,
\newblock Mod. Phys. Lett. {\bf A3}, 325 (1988).

\bibitem{Alekseev:1988tj}
A.~Y. Alekseev and S.~L. Shatashvili,
\newblock Mod. Phys. Lett. {\bf A3}, 1551 (1988).

\bibitem{'tHooft:1978hy}
G.~'t~Hooft,
\newblock Nucl. Phys. {\bf B138}, 1 (1978).

\bibitem{Polyakov:1978vu}
A.~M. Polyakov,
\newblock Phys. Lett. {\bf B72}, 477 (1978).

\bibitem{Susskind:1979up}
L.~Susskind,
\newblock Phys. Rev. {\bf D20}, 2610 (1979).

\bibitem{Arnold:1995eb}
P.~Arnold and C.-x. Zhai,
\newblock Phys. Rev. {\bf D51}, 1906 (1995), [hep-ph/9410360].

\bibitem{Braaten:1995cm}
E.~Braaten and A.~Nieto,
\newblock Phys. Rev. {\bf D51}, 6990 (1995), [hep-ph/9501375].

\bibitem{Braaten:1996jr}
E.~Braaten and A.~Nieto,
\newblock Phys. Rev. {\bf D53}, 3421 (1996), [hep-ph/9510408].

\bibitem{Kajantie:1997pd}
K.~Kajantie {\em et~al.},
\newblock Phys. Rev. Lett. {\bf 79}, 3130 (1997), [hep-ph/9708207].

\bibitem{Kajantie:2000iz}
K.~Kajantie, M.~Laine, K.~Rummukainen and Y.~Schroder,
\newblock Phys. Rev. Lett. {\bf 86}, 10 (2001), [hep-ph/0007109].

\bibitem{Hart:1999dj}
A.~Hart and O.~Philipsen,
\newblock Nucl. Phys. {\bf B572}, 243 (2000), [hep-lat/9908041].

\bibitem{Boyd:1996bx}
G.~Boyd {\em et~al.},
\newblock Nucl. Phys. {\bf B469}, 419 (1996), [hep-lat/9602007].

\bibitem{Datta:1999yu}
S.~Datta and S.~Gupta,
\newblock Phys. Lett. {\bf B471}, 382 (2000), [hep-lat/9906023].


\bibitem{Kogan:2002yr}
I.~I. Kogan, A.~Kovner and J.~G. Milhano,
\newblock JHEP {\bf 12}, 017 (2002), [hep-ph/0208053].

\bibitem{Eboli:1988fm}
O.~J.~P. Eboli, R.~Jackiw and S.-Y. Pi,
\newblock Phys. Rev. {\bf D37}, 3557 (1988).

\bibitem{Bardeen:1983st}
W.~A. Bardeen and M.~Moshe,
\newblock Phys. Rev. {\bf D28}, 1372 (1983).

\bibitem{KorthalsAltes:1994ca}
C.~P. Korthals~Altes,
\newblock Nucl. Phys. {\bf B420}, 637 (1994), [hep-th/9310195].

\bibitem{Kogut:1982ez}
J.~B. Kogut, M.~Snow and M.~Stone,
\newblock Nucl. Phys. {\bf B200}, 211 (1982).

\bibitem{Gripaios:2003ce}
B.~M. Gripaios and J.~G. Milhano,
\newblock Phys. Lett. {\bf B564}, 104 (2003), [hep-ph/0302172].

\bibitem{Gripaios:2002xb}
B.~M. Gripaios,
\newblock Phys. Rev. {\bf D67}, 025023 (2003), [hep-th/0211104].

\bibitem{Teper:1998kw}
M.~J. Teper,
\newblock hep-th/9812187.

\bibitem{schwinger} W. Brown, J.P. Garrahan, I. Kogan and A.
Kovner; \newblock Phys. Rev. {\bf D62}, 016004 (2000)
[hep-th/9912136]


\bibitem{chernsimons} I. Kogan and A. Kovner;
\newblock Phys. Rev. {\bf D53}, 4510 (1996) [hep-th/9507137]

\bibitem{ianvortices} I. Kogan;
\newblock JETP Lett. {\bf 45}, 709 (1987)


\end{thebibliography}

\end{document}